\documentclass[prb,amsmath,amsfonts,twocolumn]{revtex4}

\pdfoutput=1

\usepackage{graphicx}
\usepackage{epsfig}
\usepackage{bm}
\usepackage{dcolumn}
\usepackage{amsmath}
\usepackage{amssymb}
\usepackage{color}
\usepackage{natbib}

\newcommand{\bsub}{\begin{subequations}}
\newcommand{\esub}{\end{subequations}}

\newcommand{\rr}{\mathfrak{r}}
\newcommand{\ii}{\mathfrak{i}}

\newcommand \bea {\begin{eqnarray} }
\newcommand \eea {\end{eqnarray}}
 
\newcommand{\beg}{\begin{equation}}
\newcommand{\en}{\end{equation}}
\newcommand{\bp}{\mathbf p}
\newcommand{\bq}{\mathbf q}
\newcommand{\bk}{\mathbf k}

\newcommand \bel  {\begin{align}}
\newcommand \enl  {\end{align}}

\newcommand{\up}{\uparrow}
\newcommand{\dn}{\downarrow}
\newcommand{\dg}{^\dagger}

\newcommand{\pmat}{\begin{pmatrix}}
\newcommand{\epmat}{\end{pmatrix}}

\def\Xint#1{\mathchoice
   {\XXint\displaystyle\textstyle{#1}}%
   {\XXint\textstyle\scriptstyle{#1}}%
   {\XXint\scriptstyle\scriptscriptstyle{#1}}%
   {\XXint\scriptscriptstyle\scriptscriptstyle{#1}}%
   \!\int}
\def\XXint#1#2#3{{\setbox0=\hbox{$#1{#2#3}{\int}$}
     \vcenter{\hbox{$#2#3$}}\kern-.5\wd0}}

\def\dashint{\Xint-}

\def\8{\infty}

\def\undertext#1{\vtop{\hbox{#1}\kern 1pt \hrule}}

\def\be{\begin{equation}}
\def\ee{\end{equation}}
\def\bea{\begin{eqnarray} & &}
\def\eea{\end{eqnarray}}


\begin{document}

\title{Non-adiabatic dynamics of superfluid spin--orbit coupled degenerate Fermi gas}

\author{Maxim Dzero$^{1,2}$, Ammar A. Kirmani$^1$ and Emil A. Yuzbashyan$^3$}
\affiliation{$^1$Department of Physics, Kent State University, Kent, OH 44240 USA\\ 
$^2$Max Planck Institute for the Physics of Complex Systems, N\"{o}thnitzer str. 38, 01187 Dresden, Germany \\
$^3$Center for Materials Theory, Rutgers University, Piscataway, NJ 08854, USA}

\begin{abstract} 
We study a problem of non-adiabatic superfluid dynamics of spin--orbit coupled neutral fermions in two spatial dimensions. We focus on the two cases when the out-of-equilibrium conditions are initiated either by a sudden change of the pairing strength or the population imbalance. For the case of zero population imbalance and within the mean-field approximation, the non-adiabatic evolution of the pairing amplitude in a collisionless regime can be found exactly by employing the method of Lax vector construction. Our main finding is that the presence of the spin--orbit coupling significantly reduces the region in the parameter space where a steady state with periodically oscillating pairing amplitude is realized. For the collisionless dynamics initiated by a sudden disappearance of the population imbalance we obtain an exact expression for the steady state pairing amplitude. In the general case of quenches to a state with finite population imbalance we show that there is a region in the steady state phase diagram where at long times the pairing amplitude dynamics is  governed by the reduced number of the equations of motion in full analogy with exactly integrable case. 
\end{abstract}


\maketitle

\section{Introduction}
%
Starting with the seminal paper by Gor'kov and Rashba,\cite{Lev2001} there has been a remarkable resurgence of interest in the physical properties of the spin-orbit coupled superfluids and superconductors in the past decade.\cite{ExpSO_Spielman2011,HePRL,HePRA,HeReview,ExpSO_SaDeMelo2011,ExpSO_Zhang2012,ExpSO_Wang2012,ExpSO_Cheuk2012,ExpSO_Qu2013} This interest is largerly motivated by theoretical discovery of topological insulators and topological superconductors in which spin-orbit coupling often plays a crucial role by giving rise to the existence of robust conducting states at a system's boundaries on a background of a gapped single particle spectrum in a bulk.\cite{kanemele05,kane1,kane2,bernevig06,Sato2009PRL,Sato2010PRB,HasanKane2010,QiZhang2011} In addition, the recent discovery of the superconductivity at the interface in the oxide-based heterostructures \cite{Ohmoto2004,Reyren2007,Caviglia2008} where the inversion symmetry is naturally broken served as an additional motivation for studying both conventional and unconventional superconductivity in spin-orbit coupled systems.\cite{Joerg2015}

Of special interest are the physical properties of topological insulators and superconductors under external influences which drive these systems far-from-equilibrium. In particular, the concept of the Floquet topological insulators have been recently developed in the context of various systems external periodic driving, which leads to an inversion of the bands with different parity giving rise to metallic edge states.\cite{Tanaka2010,Kitagawa2010,Gil2011} Furthermore, several groups have generalized the idea of Floquet topological insulators to Floquet topological $s$-wave superconductors. \cite{Jiang2011,Tong2013,AlexFloquet2013,Yang2014,Poudel2014,Sacramento2015}
Most recently, it has been shown that topological Floquet superfluidity can be realized in  systems where the periodic driving is self-generated in the process of the collisionless dynamics. \cite{Matt2013PRB,Matt2014PRL,Pu2015} 

However, certain aspects of the pairing dynamics in the collisionless regime for the spin-orbit coupled systems have not been addressed yet. The aim of this paper is to close the remaining gaps in the studies of this problem. Specifically, using both exact integrability and numerical analysis we investigate how the presence of the spin--orbit coupling affects the behavior of the pairing amplitude at long times. We consider the standard protocol of inducing far from equilibrium coherent dynamics in fermionic condensates by fast switch of one of the system's parameters. In our model we allow for non-zero out-of-plane Zeeman field $h_Z$ which gives rise to the population imbalance between the fermionic atoms in two hyperfine states. Here we discuss two cases: changes in the detuning frequency of the Feshbach resonance and in the population imbalance. 

There are three relevant time scales in the problem: the first time is the perturbation time scale
$\tau_{\textrm{quench}}$ which we take to be instantaneous; the second time scale is governed by the dynamics of the Cooper pairs, $\tau_\Delta$ while the third time scale, $\tau_\varepsilon$, accounts for the relaxation due to two-particle collisions. In what follows, we consider the limit $\tau_\varepsilon\to\infty$ and analyze the dynamics of the pairing amplitude at long times $t\gg \tau_\Delta$. Importantly, we will also neglect the possibility for the pairing amplitude to become spatially inhomogeneous, which is equivalent to an assumption of having a system with a size much smaller than the superfluid coherence length. 
 
Within the mean-field theory for the reduced BCS model in the weak coupling limit, three types of steady states have been found for the quenches of the pairing strength and provided the system is initially in its ground state:\cite{Barankov2004,Dzero2006,Emil2006,Levitov2006,Levitov2007} (Regime I) gapless steady state with zero pairing amplitude $\Delta(t\to\infty)=0$; (Regime II) steady state with the constant pairing amplitude, $\Delta(t\to\infty)=\Delta_\infty$; and
(Regime III) steady state described by the undamped periodic oscillations of the pairing amplitude. Interestingly, there are no qualitative changes in the steady state phase diagram for the quenches across the $s$-wave Feschbach resonance \cite{Emil2015} as well as for the two-dimensional chiral superfluids. \cite{Matt2013PRB,Matt2014PRL} In principle, 
other steady states, such as the one in which pairing amplitude is a multiperiod function of time, can also be realized. \cite{Emil2006,Yuzbashyan2008} However, realization of these states requires that the system is initially in an excited state.
 
Perhaps the most surprising result from the earlier studies of the non-adiabatic pairing problem is the discovery of a steady state with the periodically oscillating amplitude whose analytical expression is given by the Jacobi elliptic function.\cite{Shumeiko,Barankov2004,Emil2006,Yuzbashyan2008}
Thus, the main thrust of the present work is on one hand to investigate the fate of that steady state for a condensate with equal populations and non-zero spin-orbit coupling. On the other hand, we will also investigate whether in the model with the population imbalance a system allows the realization of that steady state, i.e. the pairing amplitude is still expressed in terms of the Jacobi elliptic function, even though non-zero population imbalance precludes the full analytical description. 
 
Let us briefly summarize our results. In the first part of the paper we analyze the effect of the spin-orbit coupling on the steady state phase diagram. We find that the steady state III is realized in much narrower region of the 
phase diagram. In particular, we find that the size of the Region III is inverse proportional to the strength of the spin-orbit coupling.  Qualitatively, this effect is due to the lifting of the Kramers degeneracy by the spin-orbit coupling.
Since the total pairing amplitude is determined by the pairing in two chiral bands and the collective collisionless dynamics is reduced to a motion of two effective variables, large spin-orbit coupling effectively hinders the appearance
of the steady state with periodically oscillating amplitude. 

The remaining part of our discussion concerns the nature of the steady state for the quenches in the population imbalance. This problem has been recently studied by Y. Dong et al. [\onlinecite{Pu2015}] by solving the Bogoliubov-de Gennes equations numerically. Here we show that for the quenches to the state with equal atomic populations, the
superfluid dynamics for the pairing amplitude can, in fact, be found exactly. Specifically, we obtain an exact expression for the steady state pairing amplitude and analyze the steady state phase diagram as a function of the population imbalance in the initial state. Our results for this part are generally in agreement with those reported in Ref. [\onlinecite{Pu2015}]. Then, we continue with the discussion for the quenches to a state with finite population imbalance. For this part we had to resort to the numerical analysis of the equations of motion. Our main finding is that when the finite value of the population imbalance exceeds some critical value, we observe the dynamical reduction in the number of quantities describing the system's dynamics. In other words, the order parameter dynamics is described by the same equations of motion as in integrable case of zero population imbalance. This implies that we are able to find an analytical form for the pairing amplitude at long times, although the parameters of the solution cannot be determined exactly from the initial conditions.

In the next Section we introduce the model, briefly review its ground state properties and derive the equation of motion which describe the superfluid dynamics in terms of real functions. In Section III we analyze the possible steady states which appear as a result of quench in the pairing strength for equal atomic populations. In the first part of Section IV we discuss the steady state diagram for the quenches to the state with zero population imbalance, while in the second part
present the results of the numerical simulations for the quenches into a state with non-zero population imbalance. Secton V is followed by the concluding discussion of our results. Lastly in Appendix A and Appendix B we provide the details on the derivation of the equations of motion.

\section{Model}
Our starting point is the BCS Hamiltonian in the presence of the spin-orbit interaction in two spatial dimensions and Zeeman magnetic field term:\cite{Lev2001,Sato2009PRL,Sato2010PRB,Pu2015} 
\beg\label{Eq1}
\begin{split}
H=&\sum\limits_{\bk\alpha\beta}\left[(\xi_{\bk}\delta_{\alpha\beta}-h_Z\sigma_{\alpha\beta}^z)+\alpha_{SO}({\vec \Gamma}_\bk\cdot{\vec \sigma})\right]\hat{c}_{\bk\alpha}\dg \hat{c}_{\bk\beta}\\
&-g\sum\limits_{\bk\bk'}\hat{c}_{\bk\up}\dg\hat{c}_{-\bk\dn}\dg\hat{c}_{-\bk'\dn}c_{\bk'\up},
\end{split}
\en
where $\hat{c}_{\bk\alpha}\dg$ is a fermionic creation operator with momentum $\bk$ and spin projection $\alpha$,  $g>0$ is the pairing strength, $~{\vec \Gamma}_\bk=(k_y,-k_x)$, $\alpha_{SO}$ is the Rashba spin-orbit coupling constant, $h_Z$ is a Zeeman field which determines the degree of the population imbalance 
and $\xi_{\bk}=k^2/2-\mu$ is the single particle energies taken relative to the chemical potential $\mu$ and we set the mass of the fermions to $m=1$. In passing we note that this model, strictly speaking, is not applicable to the system of charged fermions since the orbital effects will dominate the Pauli limiting effects. 

The non-interacting part of the Hamiltonian (\ref{Eq1}) can be diagonalized, which yields a new spectrum
\beg\label{ChiralBands}
\varepsilon_{\bk\lambda}=\xi_\bk-\lambda\sqrt{h_Z^2+(\alpha_{SO}k)^2}, ~\lambda=\pm1.
\en
We can now perform the unitary transformation from the original operators to new operators, which describe the fermionic excitations in chiral bands. The analysis of the ground state properties of the model (\ref{Eq1}) can be considerably simplified 
after we employ the mean-field theory approximation in the particle-particle channel and then make a unitary transformation from the original operators $\hat{c}_{\bk\lambda}$ to a fermionic operators in chiral basis $\hat{a}_{\bk\lambda}$. The resulting mean-field Hamiltonian reads:
\beg\label{HmfChiral}
\begin{split}
{\cal H}=&\sum\limits_{\bk\lambda}\varepsilon_{\bk\lambda}\hat{a}_{\bk\lambda}\dg\hat{a}_{\bk\lambda}
-\frac{\Delta}{2}\sum\limits_{\bk\lambda}\lambda\eta_\bk^{*}\Theta_k
\hat{a}_{\bk\lambda}\dg \hat{a}_{-\bk\lambda}\dg \\&
-\frac{\overline{\Delta}}{2}\sum\limits_{\bk\lambda}\eta_\bk\widetilde{\Theta}_k\hat{a}_{-\bk\lambda}\hat{a}_{\bk\overline{\lambda}}
+\textrm{h.c.}
\end{split}
\en
Here, for convenience, we introduced the following momentum dependent functions: 
$\eta_\bk=\exp[i\tan^{-1}(k_y/k_x)]$ and
\beg\label{Thetas}
\begin{split}
\Theta_k&=\frac{\alpha_{SO}k}{R_\bk}, \quad \widetilde{\Theta}_k=\frac{h_Z}{R_\bk}, \\ R_\bk&=\sqrt{h_Z^2+(\alpha_{SO}k)^2}.
\end{split}
\en
Formally, the model (\ref{HmfChiral}) is analogous to the model discussed by Sato et al. \cite{Sato2009PRL,Sato2010PRB}. The crucial difference in our case, however, is that the pairing gap $\Delta$ is not proximity induced and instead must be determined self-consistently:
\beg\label{Delta}
\Delta=g\sum\limits_{\bk\lambda}\eta_\bk\left[\lambda\Theta_k\langle\hat{a}_{-\bk\lambda}\hat{a}_{\bk\lambda}\rangle
+\widetilde{\Theta}_k\langle\hat{a}_{-\bk\lambda}\hat{a}_{\bk\overline{\lambda}}\rangle\right].
\en
\begin{figure}[ht]
\includegraphics[width=0.9\linewidth]{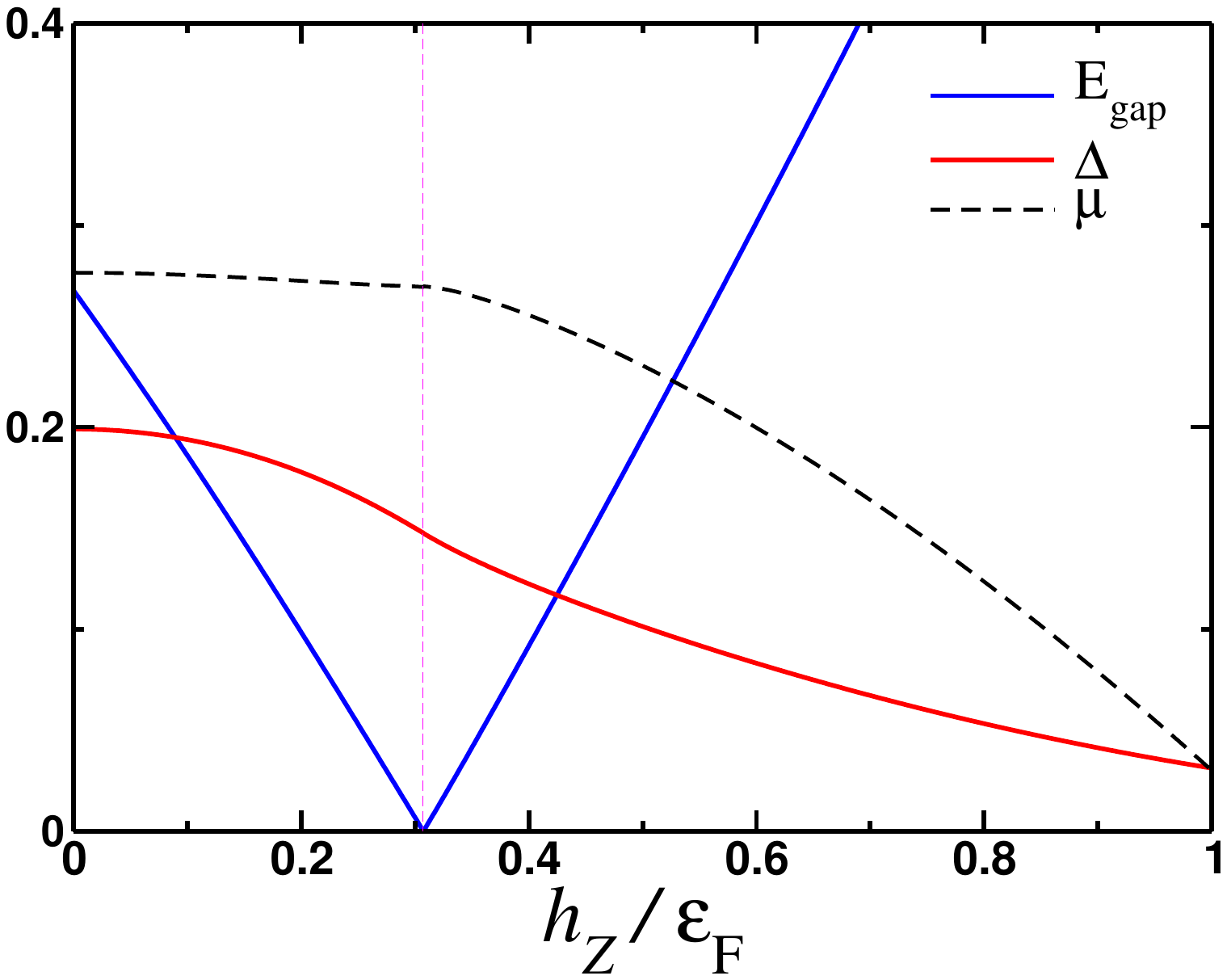}%
\caption{\label{FigGS} (Color online) Dependence of the pairing amplitude $\Delta$, chemical potential $\mu$ and spectral gap $E_{\textrm{gap}}=E_{\bk=0,\lambda=+}$ (all in arbitrary units) as a function of the Zeeman field $h_Z$ determined by the numerical solution of the self-consistency equations (\ref{DeltaGS},\ref{muGS}). Note that the superfluid becomes gapless while the order parameter remains finite at some critical value of the field $h_{Z}^{(c)}=\sqrt{\mu^2+\Delta^2}$. These results correspond to the following choice of the parameters: $n_c=0.125$, $\varepsilon_F=2\pi n_c=0.785$ and $\alpha_{SO}=0.752$.}
\end{figure}

The mean-field Hamiltonian (\ref{HmfChiral}) can be diagonalized. We find that the single particle spectrum consists of four bands $\omega_{\pm}(\bk,\lambda)=\pm E_{\bk\lambda}$ with the following dispersion
\beg\label{Eklam}
E_{\bk\lambda}=\left[\xi_\bk^2+R_\bk^2+\Delta^2-2\lambda R_\bk\sqrt{\xi_\bk^2+\widetilde{\Theta}_k^2\Delta^2}\right]^{1/2}.
\en
Before we discuss the ground state properties of the model (\ref{HmfChiral}), we first introduce the auxiliary functions which are analogous to the pseudospin variables for the BCS model. 
\subsection{Equations of motion}
In this Section we list the equations of motion (EOM) which will allow us to study the dynamics of the pairing amplitude in the collisionless regime. EOM can be obtained from the corresponding EOM for the single particle propagators, which can then be cast into the form of the EOM analogous to the Bloch equations for the magnetic moments in external magnetic field.
As a reader may have already guessed, there should be ten equations of motion overall: six equations describe the Cooper pair dynamics on each of the two chiral bands $\lambda=\pm$, while the remaining four appear as a result of non-zero Zeeman field. The details on the derivation of the equations of motion are
given in the Appendix A, so here we provide the final results. The first six equations are compactly written as follows
\beg\label{EqMot4S}
\begin{split}
\partial_t{\vec S}_{\bk\lambda}=
{\vec B}_{\bk\lambda}(t)\times{\vec S}_{\bk\lambda}(t)+{\vec m}_{\bk}(t)\times{\vec L}_{\bk\lambda}(t)
\end{split}
\en
with ${\vec B}_{\bk\lambda}=2(-\Theta_k\Delta_x,-\Theta_k\Delta_y, \varepsilon_{\bk\lambda})$ is an effective field around which ${\vec S}$ is precessing and vector ${\vec m}_k=2(-\widetilde{\Theta}_k\Delta_x,-\widetilde{\Theta}_k\Delta_y, 0)$ can be interpreted as an  "induced magnetization" since its $xy$-components vanish for $h_Z=0$. Naturally, equations (\ref{EqMot4S}) have the form of the Bloch equations for the BCS superconductor when $h_Z=0$. The first two components of ${\vec B}_{\bk\lambda}$ are determined self-consistently by
\beg\label{Eq4Delta}
\Delta_x(t)-i\Delta_y(t)=g\sum\limits_{\bk\mu}\left[\Theta_k S_{\bk\mu}^{-}(t)+\widetilde{\Theta}_k L_{\bk\mu}^{-}(t)\right]
\en
where we have adopted the usual notation $S^{\pm}=S^x\pm iS^y$. 
Equations of motion for the components of vector ${\vec L}_{\bk\lambda}(t)$ are
\beg\label{EqMot4L}
\begin{split}
{\partial_t}L_{\bk\lambda}^x=&-2\epsilon_\bk L_{\bk\lambda}^y-\widetilde{\Theta}_k\Delta_y(t)\left[S_{\bk\lambda}^z+S_{\bk\overline{\lambda}}^z\right]\\&-2\Theta_k\Delta_x(t)T_\bk, \\
{\partial_t}L_{\bk\lambda}^y=&2\epsilon_\bk L_{\bk\lambda}^x+\widetilde{\Theta}_k\Delta_x(t)\left[S_{\bk\lambda}^z+S_{\bk\overline{\lambda}}^z\right]\\&-2\Theta_k\Delta_y(t)T_\bk, \\
\partial_tL_{\bk\lambda}^z+&2\lambda R_k T_\bk(t)+\widetilde{\Theta}_k\Delta_x(t)\left[S_{\bk\lambda}^y-S_{\bk\overline{\lambda}}^y\right]\\&-\widetilde{\Theta}_k\Delta_y(t)\left[S_{\bk\lambda}^x-S_{\bk\overline{\lambda}}^y\right]=0, \\
\end{split}
\en
where $\epsilon_\bk=k^2/2$.
Note that as it follows from these equations $L_{\bk\lambda}^{x,y}=L_{\bk\overline{\lambda}}^{x,y}$ and also 
$L_{\bk\lambda}^{z}=-L_{\bk\overline{\lambda}}^{z}$.
Finally, the last equation of motion which determines the evolution of the auxiliary variable $T_\bk$ reads:
\beg\label{EqMot4T} 
\begin{split}
\partial_t T_\bk&+{\vec B}_{\bk\lambda}(t)\cdot{\vec L}_{\bk\lambda}(t)
-\frac{1}{2}\sum\limits_\lambda{\vec m}_\bk(t)\cdot{\vec S}_{\bk\lambda}(t)\\&=2\epsilon_\bk L_{\bk\lambda}^z.
\end{split}
\en
As we can immediately observe from these equations of motion, in the absence of the Zeeman field the first six equations decouple from the rest and become equivalent to the Anderson equations of motion for the pseudospins in the BCS model.\cite{BCS1957,Anderson1959} Thus, based on this observation we conclude that the evolution of ${\vec S}_{\bk\lambda}(t)$ 
can be determined exactly.\cite{Emil2005a,Emil2005b} However, for the general case of nonzero Zeeman field, one needs to resort to the numerical solution of the equations above for the dynamics initiated by a sudden change in the parameters of the model, such as pairing strength $g$, Zeeman field $h_Z$ or spin--orbit coupling $\alpha_{SO}$. In what follows we specifically study the quenches of the coupling constant and Zeeman field. 

\subsection{Initial conditions}
Let us write down the expressions for the auxiliary functions ${\vec S}_{\bk\lambda}(t)$, ${\vec L}_{\bk\lambda}(t)$ and ${T}_{\bk}(t)$ at time of a quench, $t=0$. In what follows we only focus on the case when the system is initially in its ground state. 
Then, the initial momentum distribution for these variables directly follows from the equations of motion (\ref{EqMot4S},\ref{EqMot4L},\ref{EqMot4T}). Without loss of generality, we assume that initially the superfluid order parameter is real, $\Delta_x=\Delta$, $\Delta_y=0$. Employing the relations between the single particle propagators, evaluated at equal times and auxilary functions above, for the $x$ components of ${\vec S}_{\bk\lambda}(t)$ and ${\vec L}_{\bk\lambda}(t)$ we find
\beg\label{SxLxInit}
\begin{split}
S_{\bk\lambda}^x(0)&=\frac{\Theta_k\Delta(E_{\bk\lambda}E_{\bk\overline{\lambda}}+\varepsilon_{\bk\overline{\lambda}}^2+\Delta^2)}{2E_{\bk\lambda}E_{\bk\overline{\lambda}}(E_{\bk\lambda}+E_{\bk\overline{\lambda}})}, \\
L_{\bk\lambda}^x(0)&=\frac{\widetilde{\Theta}_k\Delta[E_{\bk\lambda}E_{\bk\overline{\lambda}}+\varepsilon_{\bk\lambda}\varepsilon_{\bk\overline{\lambda}}+\Delta^2]}{2E_{\bk\lambda}E_{\bk\overline{\lambda}}(E_{\bk\lambda}+E_{\bk\overline{\lambda}})},
\end{split}
\en
while $S_{\bk\lambda}^y(0)=L_{\bk\lambda}^y(0)=T_{\bk}=0$. Reader can easily check that in the limit $h_Z=0$ we recover the expression for the Anderson pseudospin in the BCS model. Consequently, in the limit of $\alpha_{SO}=0$ we naturally 
find $S_{\bk\lambda}^x(0)=0$ while $L_{\bk\lambda}^x(0)=\varphi_\bk\Delta/2E_\bk$ with $E_\bk=\sqrt{(\epsilon_\bk-\mu)^2+\Delta^2}$ and $\varphi_\bk=E_\bk[1+\textrm{sign}(E_\bk-h_Z)]/(E_\bk+h_Z+|E_\bk-h_Z|)$. Similarly, for $S_{\bk\lambda}^z(0)$ and $L_{\bk\lambda}^z(0)$ we obtain:
\beg\label{SzLzInit} 
\begin{split}
L_{\bk\lambda}^z(0)&=\frac{\Theta_k\widetilde{\Theta}_k\Delta^2(\varepsilon_{\bk\overline{\lambda}}-\varepsilon_{\bk\lambda})}{2E_{\bk\lambda}E_{\bk\overline{\lambda}}(E_{\bk\lambda}+E_{\bk\overline{\lambda}})}, \\
S_{\bk\lambda}^z(0)&=-\frac{\varepsilon_{\bk\lambda}}{\Theta_k\Delta}S_{\bk\lambda}^x(0)
+\frac{\widetilde{\Theta}_k}{\Theta_k}L_{\bk\lambda}^z(0).
\end{split}
\en
One can easily check that in the limit $h_Z=0$ we recover the usual expression for $S_{\bk\lambda}^z(0)$ in the reduced BCS model. For the case
of finite Zeeman field and no spin--orbit coupling $L_{\bk\lambda}^z$ is zero, while 
$\Delta[S_{\bk\lambda}^z(0)]_{\alpha_{SO}=0}=-(\epsilon_\bk-\mu)[L_{\bk\lambda}^x(0)]_{\alpha_{SO}=0}$. In this case there is a similar decoupling in the equations of motion and we only need to solve six dynamics equations instead of ten. As it turns out, the pairing dynamics in this case can be found exactly.\cite{Bettel2008} Next, we discuss the ground state properties of our mean-field model. 
\subsection{Ground state}
In the absence of spin-orbit coupling superconductivity becomes energetically unfavorable when the magnitude of the Zeeman field is $h_Z>\sqrt{2}\Delta$ known as Clogston-Chandrasekar criterion.\cite{Clogston1962,Chan1962} Nonzero spin-orbit coupling, however, leads to the mixing between singlet and triplet components in the anomalous Gor'kov correlation functions \cite{Lev2001} and superconductivity extends to much higher values of the Zeeman field. 

The value of the pairing amplitude in the ground state is determined from the solution of the self-consistency equation (\ref{Eq4Delta}). Taking into account equations (\ref{SxLxInit}) above, we find
\beg\label{DeltaGS}
\begin{split}
\frac{1}{g}&=\sum\limits_{\bk\lambda}
\frac{E_{\bk\lambda}E_{\bk\overline{\lambda}}+\Delta^2+\Theta_k^2\varepsilon_{\bk\overline{\lambda}}^2+\widetilde{\Theta}_k^2
\varepsilon_{\bk{\lambda}}\varepsilon_{\bk\overline{\lambda}}}{2E_{\bk\lambda}E_{\bk\overline{\lambda}}(E_{\bk\lambda}+E_{\bk\overline{\lambda}})}.
\end{split}
\en
In addition, we need to compute the value of the chemical potential $\mu$ in the ground state. The equation for the chemical potential is obtained from the standard expression for the particle number in terms of the functions $S_{\bk\lambda}^z$. We find:
\beg\label{muGS}
\begin{split}
2n_c=\sum\limits_{\bk\lambda}&\left[\frac{1}{2}-\frac{\varepsilon_{\bk\lambda}(E_{\bk\lambda}E_{\bk\overline{\lambda}}+\varepsilon_{\bk\overline{\lambda}}^2+\Theta_k^2\Delta^2)}{2E_{\bk\lambda}E_{\bk\overline{\lambda}}(E_{\bk\lambda}+E_{\bk\overline{\lambda}})}\right.\\&\left.+\frac{\widetilde{\Theta}_k^2\Delta^2\varepsilon_{\bk\overline{\lambda}}}{2E_{\bk\lambda}E_{\bk\overline{\lambda}}(E_{\bk\lambda}+E_{\bk\overline{\lambda}})}\right],
\end{split}
\en
where we used the relation $\Theta_k^2+\widetilde{\Theta}_k^2=1$, $n_c=\varepsilon_F/2\pi$ is a particle density per spin
in two dimensions and $\varepsilon_F$ is the Fermi energy. We analyze both of these equations numerically and present the results of our analysis on Fig. \ref{FigGS}.
Perhaps the most remarkable feature of our results is the vanishing the spectral gap $E_{\textrm{gap}}=E_{\bk=0,\lambda=+}$ at some critical value of the Zeeman field $h_{Zc}=\sqrt{\mu^2+\Delta^2}$, while the pairing amplitude remains finite. This effect  is well understood: it signals a topological phase transition at which the winding number $W$ changes from $W=0$ to $W=1$ (for related discussion see e.g. Ref. [\onlinecite{Pu2015}] and references therein). The change in the winding number reflects the appearance of the Majorana gapless chiral edge modes in a sample with boundaries. 

\section{Quench of the pairing strength in the model with zero population imbalance}
In this Section we consider the pairing dynamics following the sudden change of the pairing strength for equal atomic populations, $h_Z=0$. In this case $\Theta_k=1$ and $\widetilde{\Theta}_k=0$.
We will mainly focus of the details of the steady state "phase diagram" ignoring another aspects of the problem
such as long-time asymptote of the pairing amplitude and steady state quasiparticle distribution function
due to the similarity with the corresponding problem discussed in great details by Yuzbashyan et al. [\onlinecite{Emil2015}].

\subsection{Lax vector}
Here we will introduce quantities, which we will later use to analyze the steady state dynamics of the condensate. 
The Lax vector for our problem is defined according to:
\beg\label{Lax}
\begin{split}
&\vec{\cal L}(u)=\sum\limits_{\bk\lambda}\frac{{\vec S}_{\bk\lambda}}{u-\varepsilon_{\bk\lambda}}-\frac{{\vec e}_z}{g}.
\end{split}
\en
Equation of motion for the Lax vector follows directly from the equations of motion for the pseudospins ${\vec S}_{\bk\lambda}$:
\beg\label{Eq4LaxMain}
\partial_t{\vec {\cal L}}(u)=[-2{\vec \Delta}(t)+2u{\vec e}_z]\times{\vec {\cal L}}(u).
\en
The square of the Lax vector is conserved by the evolution
\beg\label{L2}
{\vec{\cal L}}^2(u)=\frac{1}{g^2}+\sum\limits_{\bp\lambda}\left[\frac{2{\cal H}_{\bp\lambda}}{u-\varepsilon_{\bp\lambda}}+\frac{{\vec S}_{\bp\lambda}^2}{(u-\varepsilon_{\bp\lambda})^2}\right],
\en
where we have introduced 
\beg\label{Hjint}
{\cal H}_{\bp\lambda}=
\sum\limits_{\bp\lambda\not=\bq\mu}\frac{{\vec S}_{\bp\lambda}\cdot{\vec S}_{\bq\mu}}{(\varepsilon_{\bp\lambda}-\varepsilon_{\bq\mu})}-\frac{S_{\bp\lambda}^z}{g}
\en
Following the arguments of Ref. [\onlinecite{Emil2015}] we immediately conclude that the dynamics governed
by the mean-field Hamiltonian (\ref{Eq1}) with $h_Z=0$ can be determined exactly. 

Our main goal in this Section is to determine the steady state phase diagram, which we will plot in the plane
of initial and final values of the superfluid order parameters, $\Delta_{\textrm{0i}}$ and $\Delta_{\textrm{0f}}$, just
like it has been done in earlier works.\cite{Matt2013PRB,Pu2015,Emil2015}

As it has been extensively discussed in Ref. [\onlinecite{Emil2015}], in the thermodynamic limit the imaginary part of the complex roots of the spectral polynomial determine the value of the pairing amplitude in a steady state. 
Let us compute the roots of (\ref{L2}) for the initial configuration of the pseudospins. It follows:
\beg
{\cal L}_x(u,g_i)=\sum\limits_{\bk\lambda}\frac{S_{\bk\lambda}^x}{u-\varepsilon_{\bk\lambda}}=\Delta_{\textrm{0i}}{\cal L}_0(u), 
\en
where 
\beg
{\cal L}_0(u)=\sum\limits_{\bk\lambda}\frac{1}{2(u-\varepsilon_{\bk\lambda})\sqrt{(\varepsilon_{\bk\lambda}-\mu)^2+\Delta_{\textrm{0i}}^2}}.
\en
Similarly, ${\cal L}_y(u,g_i)=0$ and
\beg
{\cal L}_z(u,g_i)=-(u-\mu){\cal L}_0(u).
\en
Thus, Eq. (\ref{L2}) becomes
\beg\label{RootsEq}
[(u-\mu)^2+\Delta_{\textrm{0i}}^2]{\cal L}_0^2(u)=0.
\en
Clearly, the equation (\ref{RootsEq}) has the complex conjugates pair of roots:
\beg\label{u0}
u_{0,\pm}=\mu\pm i\Delta_{\textrm{0i}}
\en
and the imaginary part of $u_{0,\pm}$ gives the value of the pairing amplitude
We also define a spectral polynomial 
\beg\label{QN}
Q_{2N+2}(u)=g^2\prod\limits_{\bp\lambda}(u-\varepsilon_{\bp\lambda})^2\cdot{\cal L}^2(u),
\en
where $N$ is the total number of distinct single particle energy levels $\varepsilon_{\bp\lambda}$. 
Since we are considering the case when the pairing strength changes abruptly from $g_i\to g_f$, we set $g=g_f$ in Eqs. (\ref{Lax},\ref{QN}).
\subsection{Roots of the spectral polynomial and steady state diagram}
For the case when the coupling is changed instantaneously, the complex roots of Eq. (\ref{L2}) or, equivalently, the roots of the spectral polynomial (\ref{QN}) with $g=g_f$ can be obtained from
\beg\label{Roots}
\frac{\tilde{\beta}}{u-\mu\mp i\Delta_{\textrm{0i}}}+\sum\limits_{\bk\lambda}\frac{1}{2(u-\varepsilon_{\bk\lambda})\sqrt{(\varepsilon_{\bk\lambda}-\mu)^2+\Delta_{\textrm{0i}}^2}}=0,
\en
where $\tilde{\beta}=g_f^{-1}-g_i^{-1}$.
To analyze Eq. (\ref{Roots}) it is convenient 
to go from summations over momentum to the integration over energy by introducing the density of states
$\nu(\epsilon)=\nu_F$ where $\nu_F=n_c/\varepsilon_F$ and $\varepsilon_F$ is the Fermi energy, $n_c$ is a particle density per spin. 
We need to consider contribution from each chiral band separately. 

Consider $\lambda=+$ first with 
$\varepsilon_{\bk+}=k^2/2-\alpha_{SO}k$:
\beg\label{Int1}
\begin{split}
\sum\limits_{\bk}F(\varepsilon_{\bk+})&=\int\limits_{0}^\alpha\frac{kdk}{2\pi}F(\varepsilon_{\bk+})+\int\limits_{\alpha}^\infty\frac{kdk}{2\pi}F(\varepsilon_{\bk+}).
\end{split}
\en
Next, we introduce an integration variable $\epsilon={k^2}/{2}-\alpha_{SO}k$, so that:
\beg\label{kpm}
k_{\pm}(\epsilon)=\alpha_{SO}\left(1\pm\sqrt{1+\frac{2\epsilon}{\alpha_{SO}^2}}\right).
\en
For the first integral in (\ref{Int1}) we need to pick $k_{-}(\epsilon)$ while in the second integral 
we pick $k_{+}(\epsilon)$. It follows:
\beg\label{Int2}
\begin{split}
\sum\limits_{\bk}F(\varepsilon_{\bk+})&=\int\limits_{-\alpha_{SO}^2/2}^0\frac{d\epsilon}{2\pi}\frac{2\alpha_{SO}}{\sqrt{\alpha_{SO}^2+2\epsilon}}F(\epsilon)\\&+\int\limits_{0}^\infty\frac{d\epsilon}{2\pi}\left(1+\frac{\alpha_{SO}}{\sqrt{\alpha_{SO}^2+2\epsilon}}\right)F(\epsilon).
\end{split}
\en
The contribution from the chiral band $\lambda=-$ is trivial and it yields:
\beg\label{Int3}
\sum\limits_{\bk}F(\varepsilon_{\bk-})=\int\limits_{0}^\infty\frac{d\epsilon}{2\pi}\left(1-\frac{\alpha_{SO}}{\sqrt{\alpha_{SO}^2+2\epsilon}}\right)F(\epsilon).
\en
Thus, Eq. (\ref{Roots}) becomes
\beg\label{MainRoots}
\begin{split}
&\frac{{\beta}}{u-\mu\mp i\Delta_{\textrm{0i}}}+\int\limits_0^{\omega_D}\frac{d\epsilon}{2(u-\epsilon)\sqrt{(\epsilon-\mu)^2+\Delta_{\textrm{0i}}^2}}\\&+\int\limits_{-\alpha_{SO}^2/2}^0\frac{\alpha_{SO}d\epsilon}{2\sqrt{\alpha_{SO}^2+2\epsilon}(u-\epsilon)\sqrt{(\epsilon-\mu)^2+\Delta_{\textrm{0i}}^2}}=0.
\end{split}
\en
where $\beta=\tilde{\beta}/2\nu_F$ and $\omega_D$ is the bandwidth. Naturally, when $\alpha_{SO}=0$ we recover the equation for the Lax roots in the BCS model. Although in the subsequent analysis we can safely take $\omega_D\to\infty$, however, in numerical calculations we have to keep the bandwidth finite. 

We are interested in finding the values of $\beta$ for which the equation (\ref{MainRoots}) will have two pairs
of complex conjugated roots. Let us introduce the following variable:
\beg\label{u2v}
u=\mu+v\Delta_{\textrm{0i}}.
\en
The imaginary roots which determine the value of the pairing amplitude in the steady state are determined by setting
\beg
u\to u\pm i\delta.
\en
Using (\ref{u2v}) we re-write (\ref{MainRoots}) as follows:
\beg
\begin{split}
&\frac{2\beta(v\pm i)}{v^2+1}+\int\limits_{-\mu/\Delta_{\textrm{0i}}}^{\infty}\frac{d\epsilon}{(v-\epsilon)\sqrt{\epsilon^2+1}}
\\&+\sqrt{\frac{\alpha_{SO}^2}{2\Delta_{\textrm{0i}}}}\int\limits_{-(\alpha_{SO}^2+2\mu)/2\Delta_{\textrm{0i}}}^{-\mu/\Delta_i}\\ & \times
\frac{d\epsilon}{\sqrt{\epsilon+\frac{\alpha_{SO}^2+2\mu}{2\Delta_{\textrm{0i}}}}(v-\epsilon)\sqrt{\epsilon^2+1}}=0.
\end{split}
\en
Let us find the critical value of $\beta$ when the imaginary part of $v$ becomes non-zero for the first time. We have
\beg\label{TwoEqs4Roots}
\begin{split}
&\pm\frac{2\beta}{v^2+1}\mp\frac{\pi\vartheta(v\Delta_{\textrm{0i}}+\mu)}{\sqrt{v^2+1}}\\&\mp\sqrt{\frac{\alpha_{SO}^2}{2\Delta_{\textrm{0i}}}}
\frac{\pi\vartheta(-v\Delta_{\textrm{0i}}-\mu)\vartheta(2v\Delta_i+2\mu+\alpha_{SO}^2)}{\sqrt{v+\frac{\alpha_{SO}^2+2\mu}{2\Delta_{\textrm{0i}}}}\sqrt{v^2+1}}=0, \\
&\frac{2\beta v}{v^2+1}+\dashint_{-\mu/\Delta_{\textrm{0i}}}^{\infty}\frac{d\epsilon}{(v-\epsilon)\sqrt{\epsilon^2+1}}
\\ &+\sqrt{\frac{\alpha_{SO}^2}{2\Delta_{\textrm{0i}}}}
\dashint_{-(\alpha_{SO}^2+2\mu)/2\Delta_{\textrm{0i}}}^{-\mu/\Delta_{\textrm{0i}}}\frac{d\epsilon}{{\cal R}(\epsilon)}=0.
\end{split}
\en
where we introduced for brevity function
\[
{\cal R}(\epsilon)=\sqrt{\epsilon+\frac{\alpha_{SO}^2+2\mu}{2\Delta_{\textrm{0i}}}}(v-\epsilon)\sqrt{\epsilon^2+1}.
\]
Let us analyze the first equation in (\ref{TwoEqs4Roots}). Depending on the value of $v$, there are two possible solutions. First solution corresponding to the
usual BCS case:
\beg\label{vBCS}
|\beta_c|=\frac{\pi}{2}\sqrt{v^2+1}, \quad v> -\mu/\Delta_{\textrm{0i}}, 
\en
while $v$ is found by solving
\beg
\begin{split}
&\frac{\pi\textrm{sign}(\beta_c)}{\sqrt{v^2+1}}+\dashint_{-\mu/\Delta_{\textrm{0i}}}^{\infty}\frac{d\epsilon}{(v-\epsilon)\sqrt{\epsilon^2+1}}
\\&+\sqrt{\frac{\alpha_{SO}^2}{2\Delta_{\textrm{0i}}}}
\int\limits_{-(\alpha_{SO}^2+2\mu)/2\Delta_{\textrm{0i}}}^{-\mu/\Delta_{\textrm{0i}}}\frac{d\epsilon}{{\cal R}(\epsilon)}=0
\end{split}
\en
still for $v\geq -\mu/\Delta_{\textrm{0i}}.$
There is, however, another solution for $\beta_c$ given by
\beg
|\beta_c|=\frac{\pi\alpha_{SO}}{2}\frac{\sqrt{v^2+1}}{\sqrt{2\Delta_{\textrm{0i}} v+\alpha_{SO}^2+2\mu}}, ~v\leq -\mu/\Delta_{\textrm{0i}}.
\en
The value of $v$ in this case will be given by
\beg
\begin{split}
&\frac{\pi\alpha_{SO}\textrm{sign}(\beta_c)}{\sqrt{v^2+1}\sqrt{2\Delta_{\textrm{0i}} v+\alpha_{SO}^2+2\mu}}+\int\limits_{-\mu/\Delta_{\textrm{0i}}}^{\infty}\frac{d\epsilon}{(v-\epsilon)\sqrt{\epsilon^2+1}}
\\&+\sqrt{\frac{\alpha_{SO}^2}{2\Delta_{\textrm{0i}}}}
\dashint_{-(\alpha_{SO}^2+2\mu)/2\Delta_{\textrm{0i}}}^{-\mu/\Delta_{\textrm{0i}}}\frac{d\epsilon}{{\cal R}(\epsilon)}=0
\end{split}
\en
for $v\leq -\mu/\Delta_{\textrm{0i}}$.
\begin{figure}[htb]
\includegraphics[width=7.25cm,height=6cm]{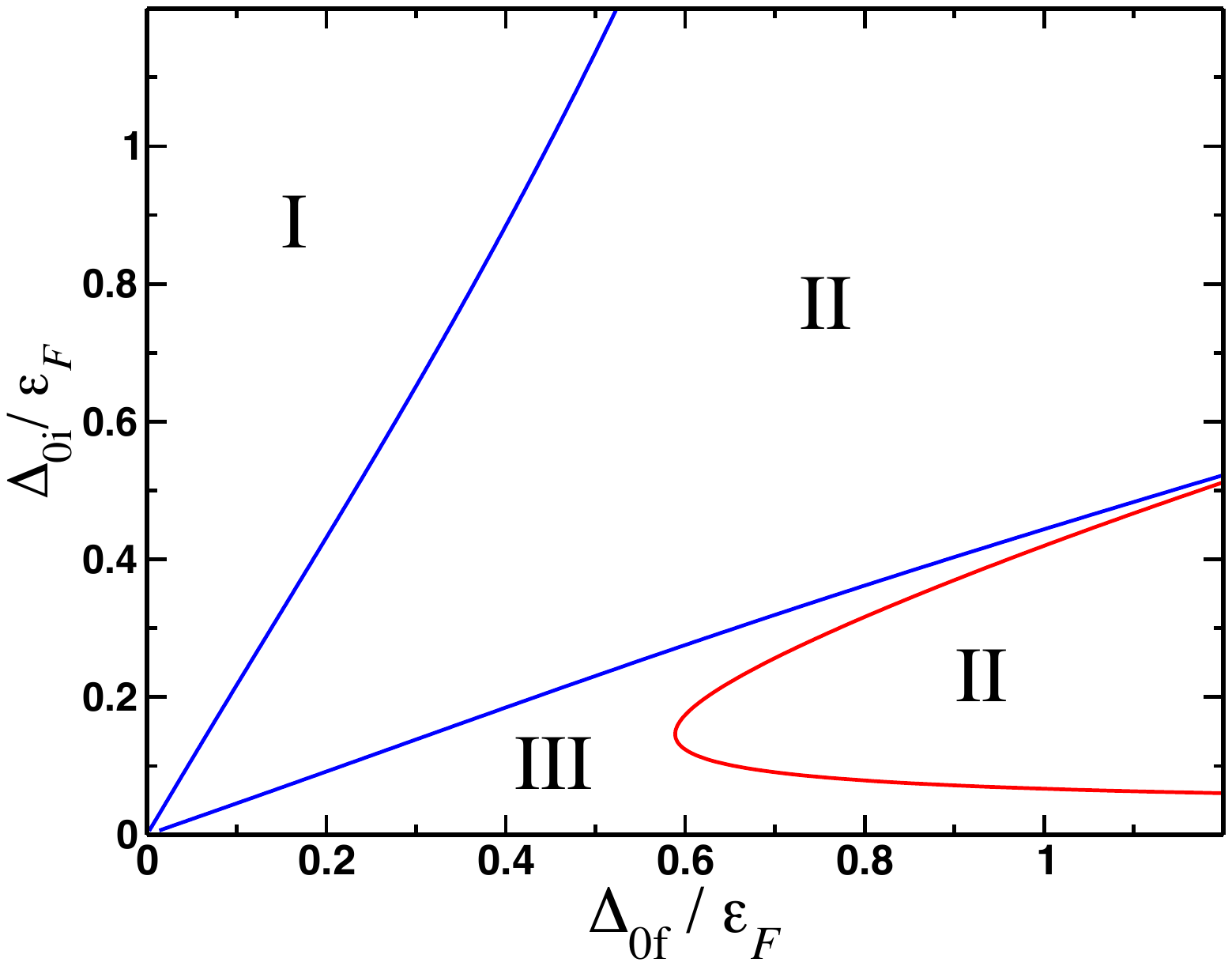}
\caption{(Color online) Steady state diagram for the case $h_Z=0$. In the region I $\Delta(t\to\infty)=0$. In the Region II $\Delta(t\to\infty)=\Delta_\infty$. Lastly, in the region III $\Delta(t)$ varies periodically with time. Parameters: $n_c=0.125$, $\alpha_{SO}=0.75$. In the limit of zero spin-orbit coupling the read line inside the Region III is absent. }
\label{FigDiaASO}
\end{figure}
The results of our analysis of equations for the critical $\beta$ above are shown on Fig. \ref{FigDiaASO}. The presence of the spin-orbit coupling leads to the appearance of the region where pairing amplitude goes to a constant (Region II) is realized inside the region where the pairing amplitude periodically varies with time (Region III). 
%
%
%
%
\section{Quench of the population imbalance}
As we have seen already, non-zero Zeeman field breaks integrability. Thus, for the quenches of the Zeeman field $h_{Zi}\to h_{Zf}$ one needs to resort to the numerical analysis of the equations of motion. The main interest in studying this particular type of quench is mainly motivated by the existence of the topological transition. The task of analyzing steady state diagram for an arbitrary values of $h_{Zf}$ has been recently accomplished by Dong et al. \cite{Pu2015}  However, as it became clear from our discussion above, for the spacial quenches such that $h_{Zf}=0$ the problem can be analyzed analytically using the same method of Lax vector construction. The only difference with the previous analysis is that an initial pseudospin distribution 
explicitly depends on $h_{Zi}$. 

\subsection{Integrable dynamics: $h_{Zf}=0$}
\begin{figure}
\includegraphics[width=0.9\linewidth]{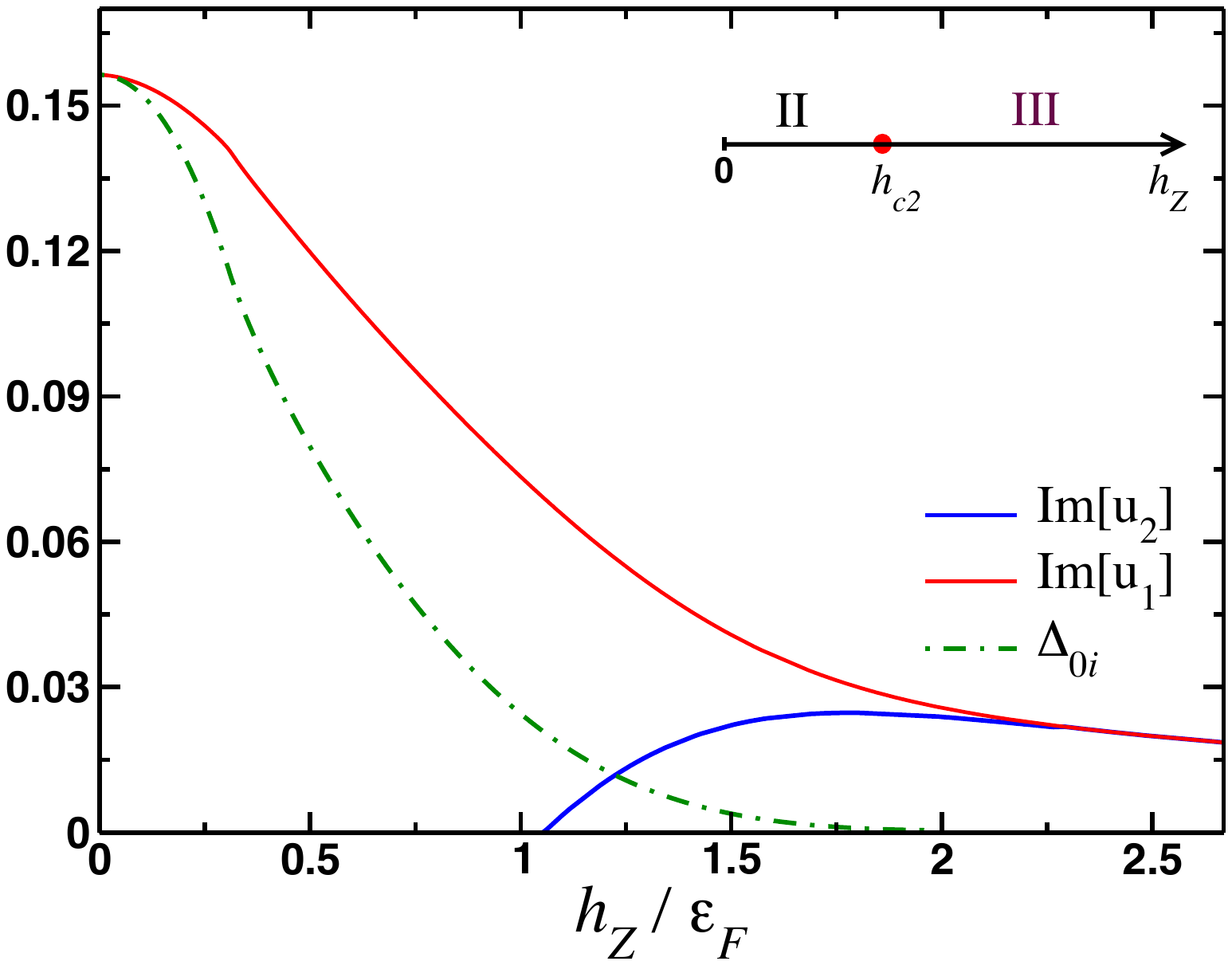}%
\caption{\label{FigQhZ} (Color online)  Imaginary parts of the roots of ${\cal L}^2(u)=0$ and superfluid order parameter $\Delta_{0i}$ in the initial state plotted as a function of initial value of the Zeeman field $h_{Zi}=h_Z$ for the quenches $h_{Zi}\to h_{Zf}=0$. Note that the imaginary parts of both roots essentially coincide with each other for the initial conditions with $\Delta_{0i}\to 0$. At $h_Z=h_{c2}$ the second complex root appears. Thus, $h_{c2}$ separates the steady states with constant and periodically oscillating superfluid order parameter. These results correspond to the following choice of the parameters: $n_c=0.125$, $\varepsilon_F=0.785$ and $\alpha_{SO}=0.752$.}
\end{figure}
\begin{figure}[ht]
\includegraphics[width=0.9\linewidth]{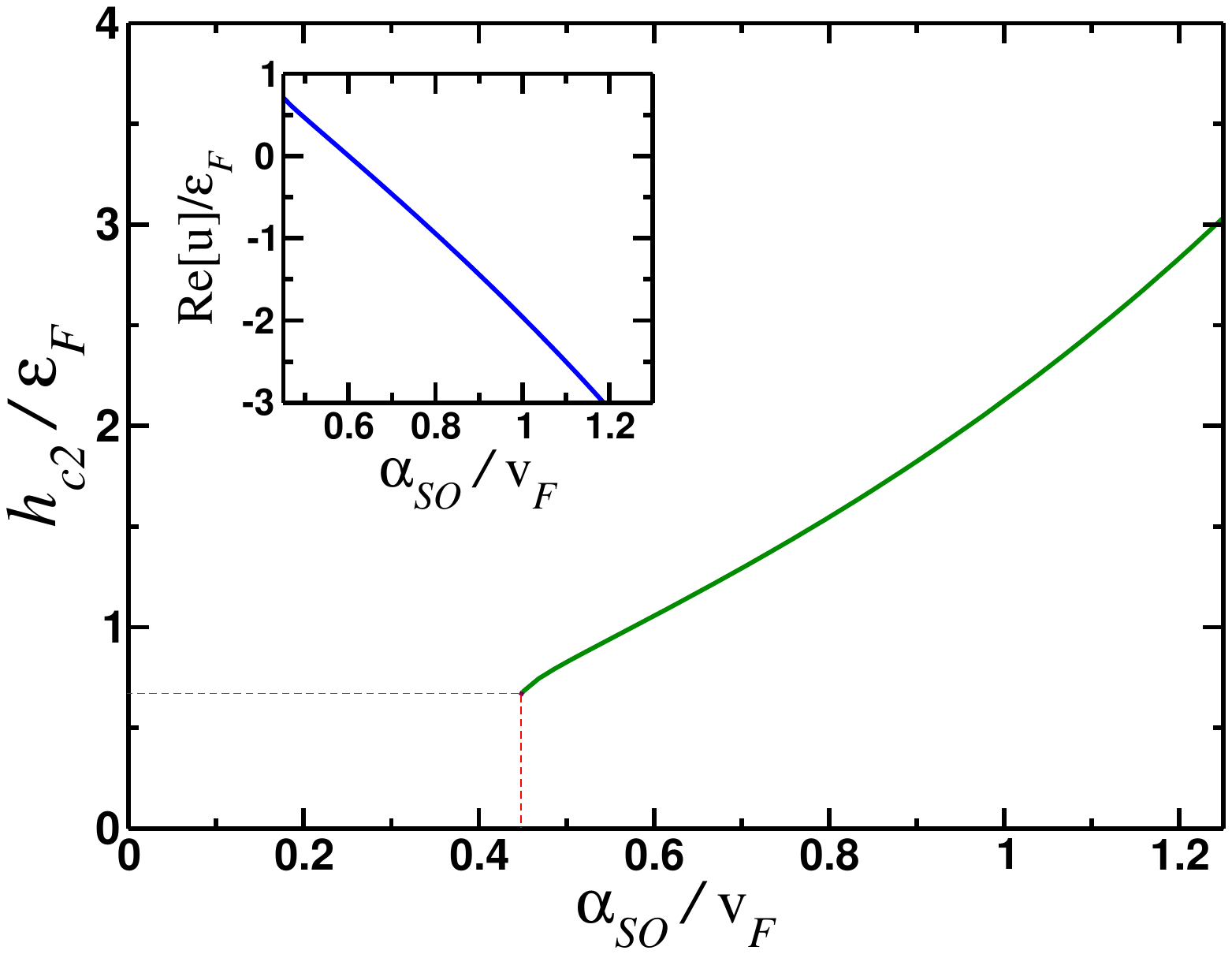}%
\caption{\label{FighZcaSO} (Color online)  The dependence of the critical Zeeman field $h_{c2}$, which separates steady state with constant and periodically oscillating pairing amplitude, on the strength of the spin--orbit coupling. Note that for the case of weak spin--orbit coupling, the pairing amplitude will always go to a constant at long times, $\Delta(t\to\infty)=\Delta_\infty$. The inset shows the dependence of the real part of the second root on $\alpha_{SO}$. These results correspond to the following choice of the parameters: $n_c=0.125$, $\varepsilon_F=0.785$.}
\end{figure}
We start with the analysis of the expression for the Lax vector (\ref{Lax}).
The expression for ${\cal L}_z(u)$ can be considerably simplified if we take into account the self-consistency equation (\ref{DeltaGS}). However, one needs to be careful, since at large fields self-consistency equation does not have a solution and we have to set $\Delta=\Delta_{0i}=0$ in (\ref{Lax}). Therefore, we have to consider two cases: in the first case $\Delta_{0i}$ in the initial state is nonzero, while in the second one $h_Z$ is large enough so that $\Delta_{0i}=0$. 

\begin{figure}[ht]
\includegraphics[width=0.8\linewidth]{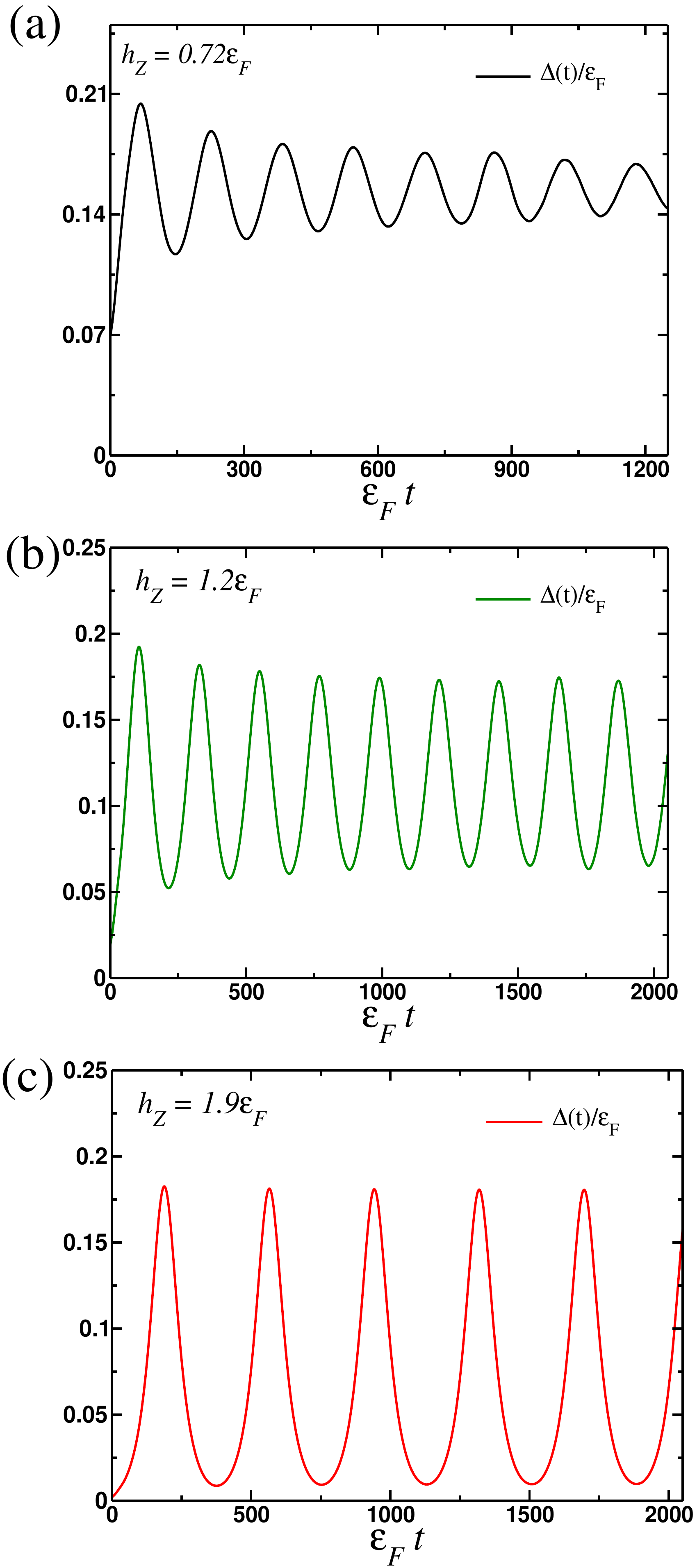}
\caption{\label{FigDynInt} (Color online) Results of the numerical solution of the equations of motion 
for the $\Delta(t)$ in the exactly integrable case when $h_{Zf}=0$. 
These results correspond to the following choice of the parameters: $n_c=0.125$, $\varepsilon_F=0.785$ and
$h_{c2}=1.02\varepsilon_F$.}
\end{figure}

We first analyze the roots for the case of finite $\Delta_{0i}$. The roots are the computed numerically from 
\beg\label{RootsFiniteDelta2}
\begin{split}
&\sum\limits_{\bk\lambda}\frac{(u-\mu+i\Theta_k\Delta)(E_{\bk\lambda}E_{\bk\overline{\lambda}}+\Delta^2+\varepsilon_{\bk\overline{\lambda}}^2)}{2(u-\mu-\varepsilon_{\bk\lambda})E_{\bk\lambda}E_{\bk\overline{\lambda}}(E_{\bk\lambda}+E_{\bk\overline{\lambda}})}\\&+
\sum\limits_{\bk\lambda}\frac{\widetilde{\Theta}_k^2\Delta^2(\varepsilon_{\bk\overline{\lambda}}-\varepsilon_{\bk\lambda})}
{2(u-\mu-\varepsilon_{\bk\lambda})E_{\bk\lambda}E_{\bk\overline{\lambda}}(E_{\bk\lambda}+E_{\bk\overline{\lambda}})}
\\&=
\sum\limits_{\bk\lambda}\frac{h_Z^2}{E_{\bk\lambda}E_{\bk\overline{\lambda}}(E_{\bk\lambda}+E_{\bk\overline{\lambda}})},
\end{split}
\en
which follows from (\ref{Lax}) and the self-consistency equation (\ref{DeltaGS}). We analyze this equation numerically and plot out results on Fig. \ref{FigQhZ}. As expected, for relatively small values of $h_{Zi}=h_Z$ there is only one complex root, which means that the steady state order parameter asymptotes to a constant. As the value of the field is increased further, it reaches $h_{c2}$ where the second complex conjugated root appears. For quenches of Zeeman field with $h_Z>h_{c2}$
pairing amplitude periodically oscillates in time. Our results confirm those found from the numerical simulations. \cite{Pu2015} Indeed, on Fig. \ref{FigDynInt} we show $\Delta(t)$ found by numerically solving the equations of motions for various values of $h_Z$ and
it is clearly in agreement with our analysis of the Lax roots, Fig. \ref{FigQhZ}. 

In Fig. \ref{FighZcaSO} we also plot the dependence of $h_{c2}$ on $\alpha_{SO}$, which we determine by setting $u=u_0+i\delta$ in (\ref{RootsFiniteDelta2}) and solving them together with Eqs. (\ref{DeltaGS},\ref{muGS}). As one may have expected,
$h_{c2}\propto\alpha_{SO}p_F$. Furthermore, the fact that we do not find a solution for small $\alpha_{SO}$ is in qualitative agreement with an observation that the steady state with oscillating pairing amplitude generally appears for moderate to strong quenches. 

Next, we would like to show that no more complex roots appear at large fields when $\Delta_{0i}$ is infinitesimally small. First, let us consider the case when the self-consistency equation (\ref{DeltaGS}) does not have a solution and, as before, we set $u=u_0+i\delta$. Then, in the equation for the Lax roots ${\cal L}_z(u)=\pm i{\cal L}_x(u)$ we can consider the real and imaginary parts separately. Equation for the imaginary part is satisfied only if $u_0<-h_Z$, while the equation of the real part reads
\beg\label{Eq4u0}
\frac{2}{g}+\sum\limits_{\bk\lambda}P\left(\frac{\textrm{sign}(\varepsilon_{\bk\lambda})}{u_0-\mu-\varepsilon_{\bk\lambda}}\right)=0,
\en
where $P$ stands for the principal value. We have analyzed this equation and did not find a value of coupling $g$ consistent with the zero value of the superfluid gap. We reach the same conclusion from the analysis of Eq. (\ref{RootsFiniteDelta2})
for the case when $\Delta_{0i}$ is small enough so it can be neglected. To summarize, we find that for the quenches of the Zeeman field from some finite value $h_{Zi}$ to zero, there are only two steady states possible at long times: in the first one
$\Delta(t)$ asymptotes to a constant, while in the second one $\Delta(t)$ continues to oscillate periodically. 

\subsection{Analytical solution for the pairing amplitude}
In this subsection we derive the analytic expressions for the pairing amplitude $\Delta(t)$ in a steady state. 
Our discussion here follows closely the related discussion in Refs. [\onlinecite{Matt2013PRB,Emil2015}].

Steady states with constant and periodically oscillating pairing amplitude can be described analytically by constructing the Lax vector for an effective $m$-pseudospin system. The Lax reduction procedure states that at long times the dynamics of a superfluid is governed by a dynamics of only few generalized pseudospin variables, which we will denote by ${\vec \sigma}_j$. ~\cite{Matt2013PRB,Emil2015} The Lax vector describing the reduced solution is
\beg\label{LaxRed2}
\begin{split}
&{\vec{\cal L}}_{\textrm{red}}(u)=\left(1+\sum\limits_{\bp\lambda}\frac{d_{\bp\lambda}}{u-\varepsilon_{\bp\lambda}}\right){\vec{\cal L}}_m(u), \\
&{\vec{\cal L}}_m(u)=\left(\sum\limits_{j=1}^{m}\frac{{\vec \sigma}_j}{u-\epsilon_j}-\frac{{\vec e}_z}{g}\right).
\end{split}
\en
Here ${\vec{\cal L}}_m(u)$ is a Lax vector for a reduced system, time-dependent vectors ${\vec \sigma}_j$ and parameters $d_{\bp\lambda}, \epsilon_j$ need to be determined. As it can be easily seen, vectors ${\vec \sigma}_j$ satisfy the 
same equations of motion as original pseudospins ${\vec S}_{\bp\lambda}$. 
The parameters of the reduced Lax vector are chosen such that 
\beg\label{MainReduce}
{\vec {\cal L}}(u)={\vec {\cal L}}_{\textrm{red}}(u).
\en
Therefore, the equation of motion for vector ${\vec {\cal L}}_{\textrm{red}}(u)$ is the same as the one for ${\vec {\cal L}}(u)$:
\beg\label{Eq4Laxred}
\partial_t{\vec {\cal L}}_{\textrm{red}}(u)=[-2{\vec \Delta}(t)+2u{\vec e}_z]\times{\vec {\cal L}}_{\textrm{red}}(u),
\en
where we use ${\vec \Delta}=(\Delta_x,\Delta_y)$ for brevity.
By matching the residues at $u=\varepsilon_{\bp\lambda}$ and at $u=\epsilon_j$ we find the following set of relations:
\beg\label{residues}
\begin{split}
&\sum\limits_{\bp\lambda}\frac{d_{\bp\lambda}}{\epsilon_j-\varepsilon_{\bp\lambda}}=-1, \quad j=1,...,m, \\
&d_{\bp\lambda}{\vec{\cal L}_m(\varepsilon_{\bp\lambda})}={\vec S}_{\bp\lambda}.
\end{split}
\en

In the thermodynamic limit it is possible to find the reduced solutions that have the same integrals of motion as the solutions for the quenched dynamics, i.e. they have the same ${\cal L}^2(u)$. Thus, equation (\ref{LaxRed2}) becomes 
\beg\label{LaxRed3}
1+\sum\limits_\lambda\int\frac{\nu_\lambda(\epsilon_\lambda)d_\lambda(\epsilon_\lambda)d\epsilon_\lambda}{u-\epsilon_\lambda}=
-\zeta(u)\sqrt{\frac{{\vec{\cal L}^2(u)}}{\vec{\cal L}_m^2(u)}},
\en
where $\zeta(u)=\pm1$, $\nu_\lambda(\epsilon)$ is the density of states for the chiral band $\lambda$ and ${\vec{\cal L}^2(u)}$ is determined by the initial conditions. By setting $u=\varepsilon\pm i\delta$ we can immediately determine $d_\lambda(\epsilon)$
\beg\label{deps}
d_\lambda(\varepsilon)=\frac{i\zeta(\varepsilon)}{2\pi\nu_\lambda(\varepsilon)}\left(\frac{\sqrt{{\cal L}^2(\varepsilon_-)}}{\sqrt{{\cal L}_m^2(\varepsilon_-)}}-\frac{\sqrt{{\cal L}^2(\varepsilon_+)}}{\sqrt{{\cal L}_m^2(\varepsilon_+)}}\right)
\en
with $\varepsilon_\pm=\varepsilon\pm i\delta$. In what follows, we will derive the explicit expressions to determine parameters ${\vec \sigma}_j$ and $\eta_j$ ($j\geq1$), which define ${\vec {\cal L}}_m(u)$, in terms of the complex roots of ${\cal L}^2(u)$.

\paragraph{$m=1$ solution.}
This is the case of the one-spin solution. The expression for ${\vec {\cal L}}_{m=1}$ reads: 
\beg\label{Lm1}
{\vec {\cal L}_{m=1}}(u)=\frac{{\vec \sigma}_1}{u-\epsilon_1}-\frac{{\vec e}_z}{g}.
\en
The relation between ${\vec \sigma}_1$ and $\Delta$ follows directly from the self-consistency equation (\ref{Eq4Delta}) and Eq. (\ref{residues}):
\beg\label{Delta1}
{\vec \Delta}(t)=g{\vec \sigma}_1(t),
\en
which also implies that $\sigma_1^z$ remains constant. 
Using the equation of motion for the Lax vector (\ref{Eq4Laxred}) together with (\ref{Delta1}) we can now solve for $\Delta(t)$:
\beg\label{Sol4Delta}
\Delta(t)=\Delta_\infty e^{2i\mu_\infty t-i\varphi_0},
\en
with $\mu_\infty=\epsilon_1+g\sigma_1^z$ and $\varphi_0$ is an integration constant. Parameters $\{\Delta_\infty, \mu_\infty\}$ can 
be expressed in terms of the roots for the square of the reduced Lax vector. Recall, that in the thermodynamic limit these roots are the same as the roots of ${\vec {\cal L}}^2(u)=0$ by construction. As we have seen in the previous section, when $h_{Zi}\leq h_{c2}$ there is
only one pair of complex conjugated roots, which we denote $u_\pm=u_{1\rr}\pm iu_{1\ii}$. Taking the square of the both parts in Eq. (\ref{Lm1}) and regrouping the terms in the right-hand-side yields:
\beg
u_{1\rr}=\mu_\infty, \quad u_{1\ii}=\Delta_\infty. 
\en
Thus, in agreement with earlier results we find, that the imaginary root of ${\cal L}^2(u)=0$ determines the value of the pairing amplitude at long times. 
\begin{figure}[ht]
\includegraphics[width=0.9\linewidth]{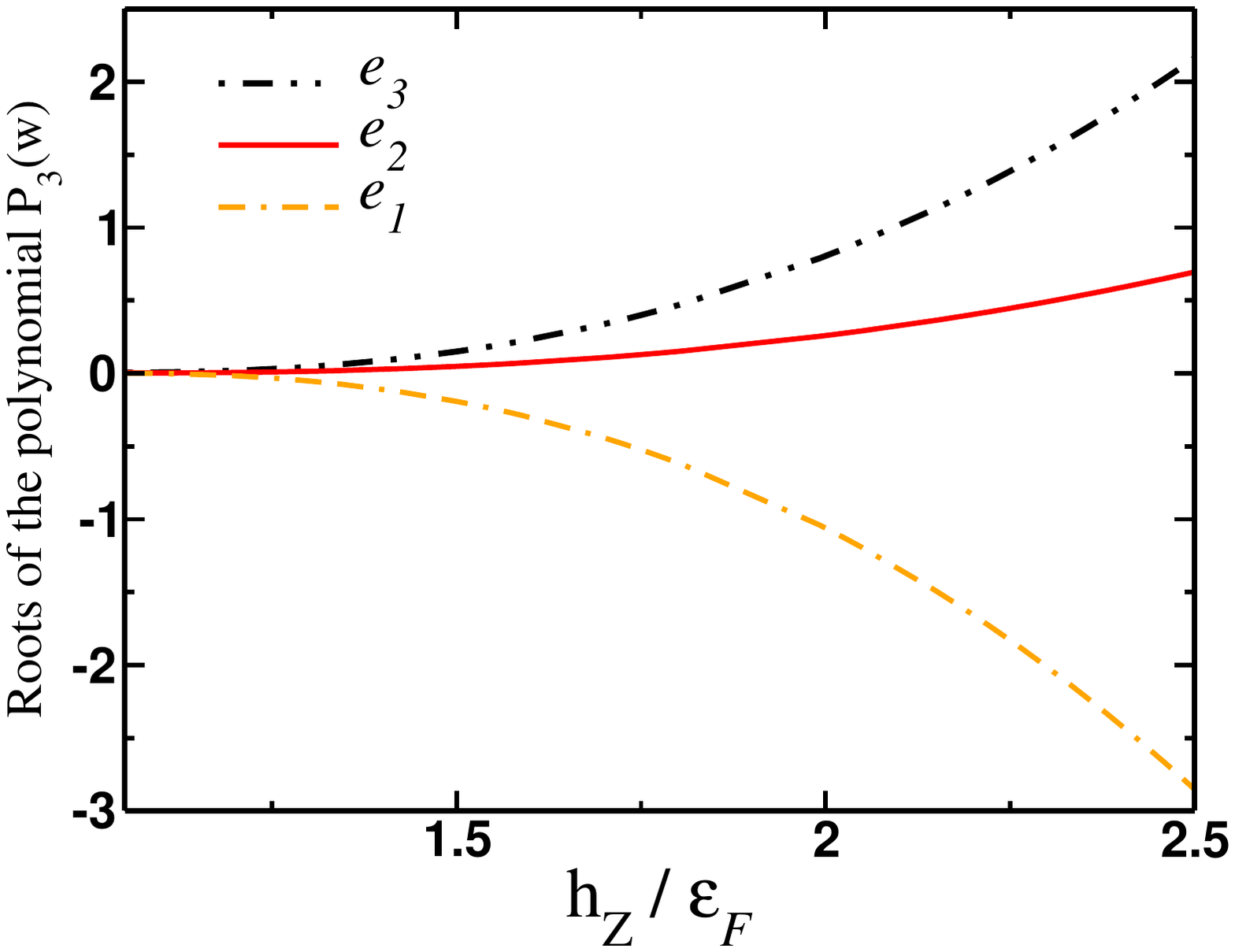}%
\caption{\label{Rootse123} (Color online) Dependence of the roots $e_1,e_2$ and $e_3$ of the qubic polynomial $P_3(w)$, Eq. (\ref{Pw}) on the value of the imbalance $h_Z$. $n_c=0.125$, $\varepsilon_F=0.785$.}
\end{figure}

\paragraph{$m=2$ solution.} This is the case of the two-spin solution with the reduced Lax vector of the form
\beg\label{Lm2}
{\vec {\cal L}_{m=2}}(u)=\frac{{\vec \sigma}_1}{u-\epsilon_1}+\frac{{\vec \sigma}_2}{u-\epsilon_2}-\frac{{\vec e}_z}{g}
\en
and
\beg\label{Dm2}
{\vec \Delta}=g\cdot({\vec \sigma}_1+{\vec \sigma}_2), \quad \Delta_x(t)-i\Delta_y(t)=\Omega e^{-i\Phi},
\en
where in the second expression $\Omega=|{\vec \Delta}|$ and $\Phi$ is the phase of the pairing amplitude. The dynamics of the variables ${\vec \sigma}_{1,2}$ is governed by the following two-spin Hamiltonian:
\beg\label{Hm2}
{H}_{m=2}=2(\epsilon_1\sigma_1^z+\epsilon_2\sigma_2^z)-2{\vec \Delta}\cdot{\vec \sigma}.
\en
The $z$-component of ${\vec \sigma}={\vec \sigma}_1+{\vec \sigma}_2$ is conserved by evolution governed by the reduced Hamiltonian (\ref{Hm2}), which reflects the total particle conservation. In addition, the total energy ${\cal E}$ must be conserved by the evolution. Given the self-consistency condition (\ref{Dm2}) for the reduced Hamiltonian, it follows that ${\cal E}$ is conserved provided the terms containing $\Delta(t)$ drop out from (\ref{Hm2}). In turn, this is only possible for $\sigma_{1,2}^z\propto{\vec \sigma}^2$. Therefore, we write:\cite{Matt2013PRB,Emil2015}
\beg\label{CritTorus}
\sigma_1^z=\frac{a_1}{g}\Omega^2+b_1, \quad \sigma_2^z=\frac{a_2}{g}\Omega^2+b_2,
\en
where coefficients $a_{1,2}$ and $b_{1,2}$ satisfy 
\beg\label{Constraintab}
\begin{split}
&\epsilon_1a_1+\epsilon_2a_2=\frac{1}{2}, \quad 2(\epsilon_1b_1+\epsilon_2b_2)={\cal E}, \\
&a_1=-a_2, \quad b_1+b_2=\sigma^z.
\end{split}
\en
Importantly, by a virtue of the second equation (\ref{residues}) we obtain the following ansatz for the original variables:
\beg\label{Ansatz4Spl}
S_{\bp\lambda}^z=a_{\bp\lambda}\Omega^2+b_{\bp\lambda}.
\en
Furthermore, equations of motion for the two remaining components of ${\vec S}_{\bp\lambda}$ - Eq. (\ref{EqMot4S}) with ${\vec m}=0$ and $\Theta_k=1$ - yields:
\beg\label{Sbpl}
S_{\bp\lambda}^{-}e^{i\Phi}-S_{\bp\lambda}^{+}e^{-i\Phi}=2ia_{\bp\lambda}\dot\Omega,
\en
where we use the notation $S_{\bp\lambda}^{\pm}=S_{\bp\lambda}^{x}\pm iS_{\bp\lambda}^y$. After a series of algebraic manipulations identical to the ones in Refs. [\onlinecite{Matt2013PRB,Emil2015}], we find the following equation for $\Omega$:
\beg\label{Eq4Omega}
\begin{split}
\dot{\Omega}^2+\Omega^4&+\left(\frac{2b_{\bp\lambda}}{a_{\bp\lambda}}+4\varepsilon_{\bp\lambda}^2\right)\Omega^2-4A\varepsilon_{\bp\lambda}\Omega \\ &+A^2+\frac{b_{\bp\lambda}^2-S_{\bp\lambda}^2}{a_{\bp\lambda}^2}=0,
\end{split}
\en
where $A$ is a function of $\Omega$ given by
\beg\label{A}
A=2\mu_A\Omega+\frac{\kappa_A}{\Omega}
\en
and $\mu_{A}$, $\kappa_A$ are arbitrary real constants. Since the same equation for $\Omega$ is found by considering the equations of motion for the variables ${\vec \sigma}_{1,2}$ we conclude that the coefficients in Eq. (\ref{Eq4Omega}) must be independent of $\bp$ and $\lambda$:
\beg\label{constants}
\begin{split}
&\frac{b_{\bp\lambda}}{a_{\bp\lambda}}+2(\varepsilon_{\bp\lambda}-\mu_A)^2=2\rho, \\
&\frac{b_{\bp\lambda}^2-S_{\bp\lambda}^2}{a_{\bp\lambda}^2}-4\kappa_A(\varepsilon_{\bp\lambda}-\mu_A)=4\chi.
\end{split}
\en
Thus the differential equation for $\Omega(t)$ becomes
\beg\label{dotOmega}
\dot{\Omega}^2+\Omega^4+4\rho\Omega^2+\frac{\kappa_A^2}{\Omega^2}+4\chi=0.
\en
Solution of this equation is:\cite{Emil2015}
\beg\label{OmSol}
\Omega=\sqrt{\Lambda^2+e_1}, \quad \Lambda=\Delta_{+}\textrm{dn}[\Delta_{+}(t-t_0),k'],
\en
where $\textrm{dn}$ is the Jacobi elliptic function, $k'={\Delta_{-}}/{\Delta_{+}}$, $\Delta_{-}^2=e_2-e_1$, $\Delta_{+}^2=e_3-e_1$ and the parameters $e_{1,2,3}$ are the real roots
of the qubic polynomial
\beg\label{Pw}
P_3(w)=w^3+4\rho w^2+4\chi w + \kappa_A^2.
\en

The last step is to match the coefficients in the polynomial (\ref{Pw}) with the values of the complex conjugated roots appearing for $h_{Zi}>h_{c2}$, Fig. \ref{FigQhZ}. To do that, we will employ the relation (\ref{residues}). First we solve Eqs. (\ref{constants}) for $a_{\bp\lambda}$, $b_{\bp\lambda}$. We find
\beg\label{aplbpl}
\begin{split}
&a_{\bp\lambda}=-\frac{S_{\bp\lambda}}{2\sqrt{[(\varepsilon_{\bp\lambda}-\mu_A)^2-\rho]^2-\kappa_A(\varepsilon_{\bp\lambda}-\mu_A)-\chi}}, \\
&b_{\bp\lambda}=\frac{[(\varepsilon_{\bp\lambda}-\mu_A)^2-\rho]S_{\bp\lambda}}{\sqrt{[(\varepsilon_{\bp\lambda}-\mu_A)^2-\rho]^2-\kappa_A(\varepsilon_{\bp\lambda}-\mu_A)-\chi}}.
\end{split}
\en
Similarly, the coefficients $a_{1,2}$, $b_{1,2}$ of the reduced solution (\ref{CritTorus}) are found using the conservation laws (\ref{Constraintab}):
\beg\label{a1b1}
a_{1,2}=\pm\frac{1}{2(\epsilon_1-\epsilon_2)}, \quad b_{1,2}=\pm\frac{{\cal E}-\epsilon_{2,1}\sigma^z}{\epsilon_1-\epsilon_2}.
\en 
Using these expressions, let us match the pre-factors in front of $\Omega^2$ after we use Eqs. (\ref{CritTorus},\ref{Ansatz4Spl}) together with (\ref{aplbpl},\ref{a1b1}) in second equation in (\ref{residues}) for the $z$-components of ${\cal L}_m$ and ${\vec S}_{\bp\lambda}$. We find:
\beg\label{res2}
\begin{split}
\frac{d_{\bp\lambda}}{g}=\frac{(\varepsilon_{\bp\lambda}-\epsilon_1)(\varepsilon_{\bp\lambda}-\epsilon_2)S_{\bp\lambda}}{\sqrt{[(\varepsilon_{\bp\lambda}-\mu_A)^2-\rho]^2-\kappa_A(\varepsilon_{\bp\lambda}-\mu_A)-\chi}}.
\end{split}
\en
On the other hand
\beg\label{Lm2}
d_{\bp\lambda}=\frac{S_{\bp\lambda}}{\sqrt{{\cal L}_{m=2}^2(\varepsilon_{\bp\lambda})}}.
\en
Introducing the spectral polynomial $Q_4(u)$ similar to (\ref{QN}):
\beg\label{Q4}
Q_4(u)=g^2(u-\epsilon_1)^2(u-\epsilon_2)^2\cdot{\vec {\cal L}}_{m=2}^2(u).
\en 
If we now compare (\ref{Q4}) with (\ref{res2}) we immediately identify $Q_4(u)$ with
\beg\label{Q4U}
Q_4(u)=[(u-\mu_A)^2-\rho]^2-\kappa_A(u-\mu_A)-\chi.
\en
Furthermore, since in the thermodynamic limit the complex roots of $Q_4(u)$ must match the complex roots of ${\cal L}^2(u)$, we can express
all the parameters (\ref{Q4U}) in terms of two pairs of complex conjugated roots $u_{1,2}=u_{1,2\rr}+iu_{1,2\ii}$:
\beg\label{coeffu12}
\begin{split}
\mu_A&=\frac{u_{1\rr}+u_{2\rr}}{2}, \\
\rho&=3\mu_A^2-2u_{1\rr}u_{2\rr}-\frac{u_{1\rr}^2+u_{1\ii}^2+u_{2\rr}^2+u_{2\ii}^2}{2}, \\
\kappa_A&=2u_{1\rr}(u_{2\rr}^2+u_{2\ii}^2)+2u_{2\rr}(u_{1\rr}^2+u_{1\ii}^2)\\ &+4\mu_A(\rho-\mu_A^2), \\
\chi&=\kappa_A\mu_A+(\mu_A^2-\rho)^2\\&-(u_{1\rr}^2+u_{1\ii}^2)(u_{2\rr}^2+u_{2\ii}^2).
\end{split}
\en

We plot the dependence of the roots of $P_3(w)$ (\ref{Pw}) on Fig. \ref{Rootse123}. Note that $e_1,e_2$ and $e_3$ are small for $h_Z\sim h_{c2}$. It was noted in Ref. \onlinecite{Emil2015} for the quenches of the detuning frequency across the Feshbach resonance, the value of $e_1$ serves as a measure of the deviation from the weak coupling limit when
$|e_1|\ll 1$. 
\begin{figure}[ht]
\includegraphics[width=0.9\linewidth]{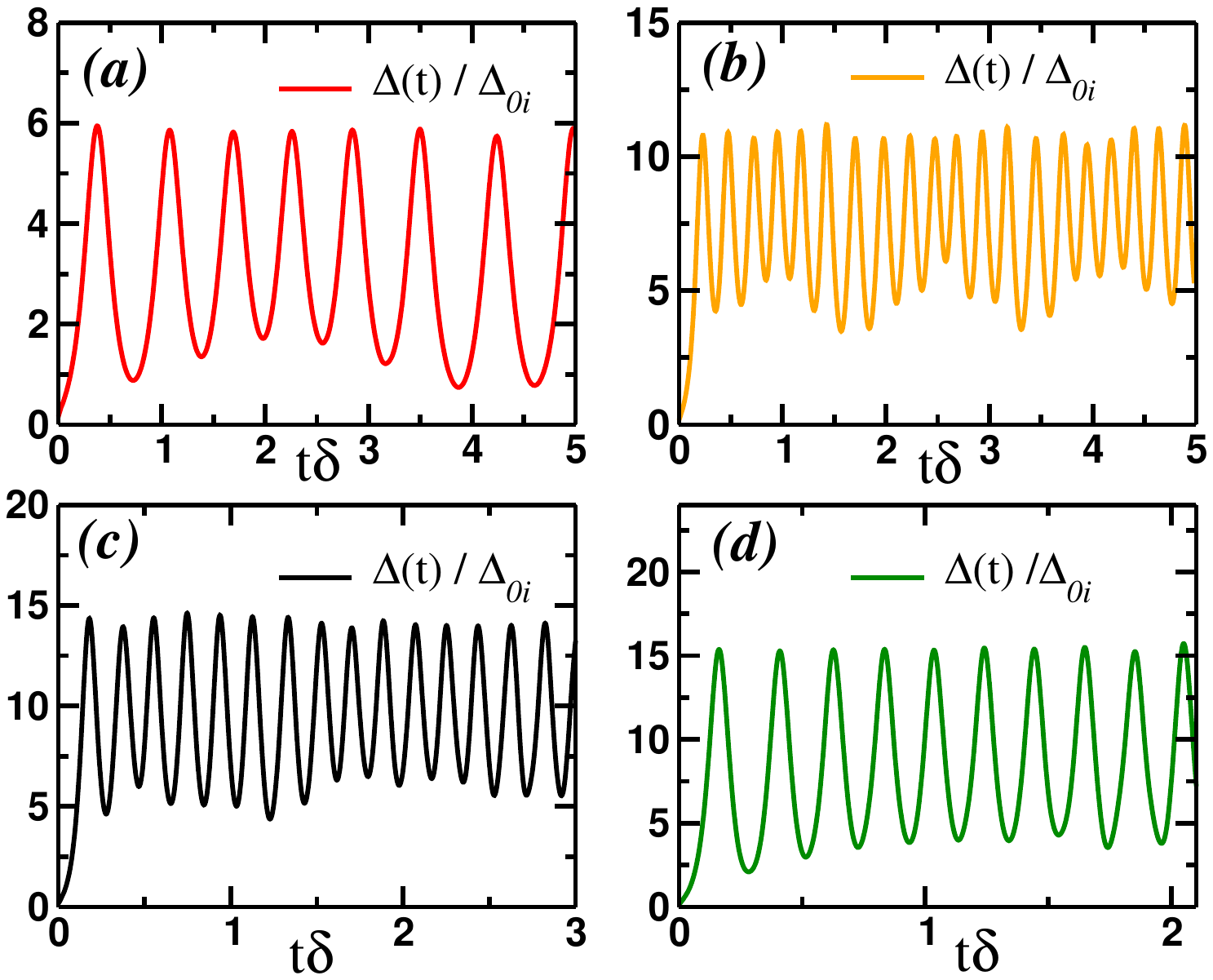}%
\caption{\label{FigDynNonint} (Color online) Results of the numerical solution of the equations of motion 
for the $\Delta(t)$ in the general, i.e. non-integrable, case $h_{Zf}\not=0$: (a) $h_{Zf}=0.9\varepsilon_F$; (b) $h_{Zf}=0.5\varepsilon_F$;
(c) $h_{Zf}=0.25\varepsilon_F$ and (d) $h_{Zf}=0.1\varepsilon_F$.
The values of the remaining parameters are: $h_{Zi}=1.85\varepsilon_F$, $n_c=0.125$, $\varepsilon_F=0.785$.}
\end{figure}
\begin{figure}[ht]
\includegraphics[width=0.9\linewidth]{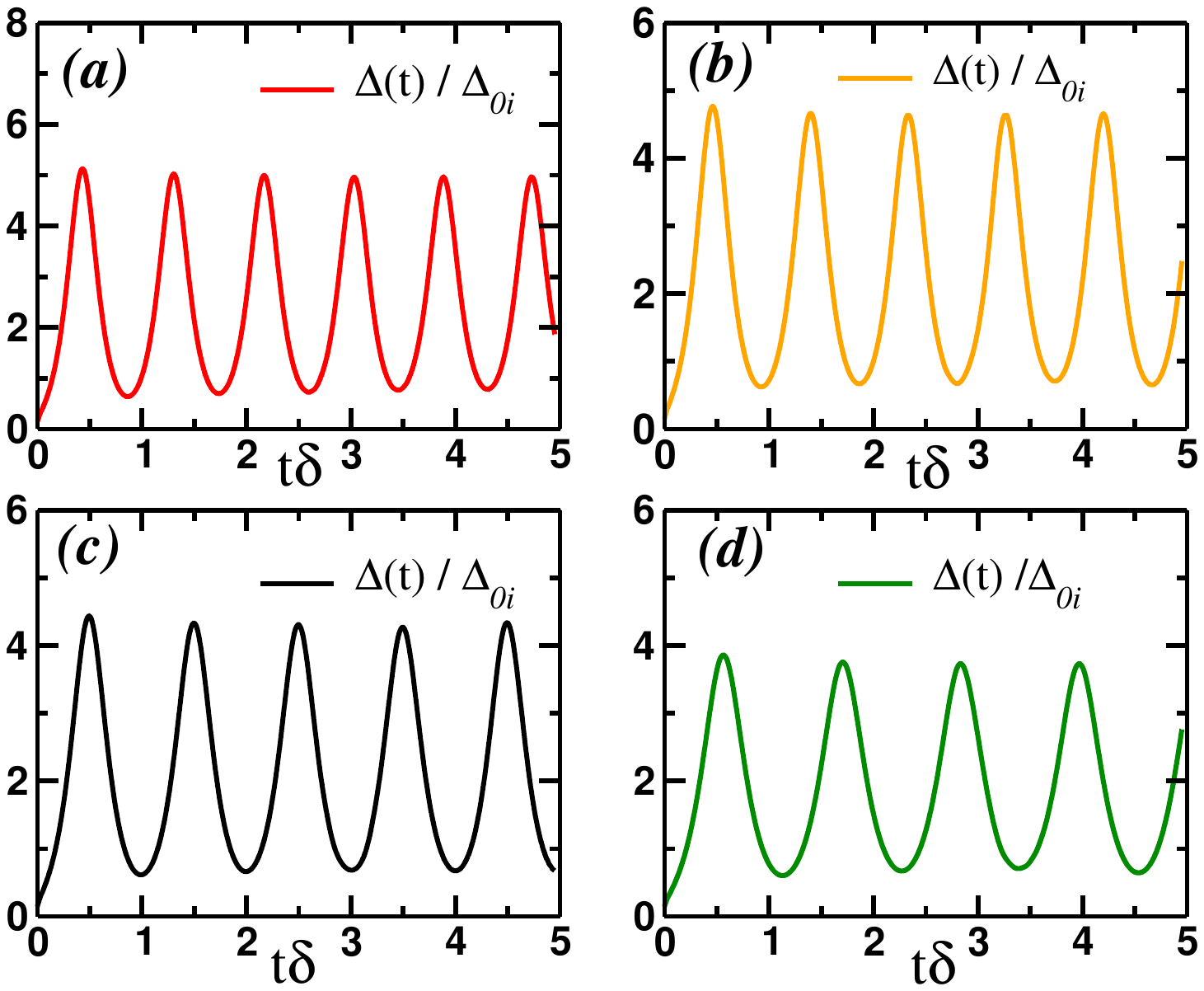}%
\caption{\label{FigDynint} (Color online) Same as Fig. \ref{FigDynNonint} with (a) $h_{Zf}=1.25\varepsilon_F$; (b) $h_{Zf}=1.15\varepsilon_F$;
(c) $h_{Zf}=1.1\varepsilon_F$ and (d) $h_{Zf}=0.95\varepsilon_F$.}
\end{figure}

To summarize, the equations (\ref{coeffu12}) together with (\ref{dotOmega}) provide exact description of the order parameter dynamics in a steady state determined by the two pairs of the complex conjugated roots of the spectral polynomial. In particular, the pairing amplitude is given by 
\beg\label{ModDelta}
|\Delta(t)|=\sqrt{e_1+\Delta_{+}^2\textrm{dn}^2[\Delta_{+}(t-t_0),k']},
\en
where the parameters entering into this expression are given above, (\ref{OmSol}). Note that parameter $e_1$ is close to zero only when $h_Z\sim h_{c2}$. It is somewhat surprising to find that $|\Delta(t)|$ is described by the weak-coupling 
solution\cite{Emil2015} 
\beg
|\Delta(t)|\propto\textrm{dn}[\Delta_{+}(t-t_0),k']
\en
only at lower fields. 

\subsection{Pairing amplitude dynamics with finite population imbalance}
Here we will discuss the dynamics initiated by the quenches of the Zeeman field, so that $h_{Zf}\not=0$. Since the dynamics governed by the Hamiltonian (\ref{Eq1}) is non-integrable, we have to resort to the numerical analysis of the equations of motion (\ref{EqMot4S},\ref{EqMot4L},\ref{EqMot4T}). Our main motivation for this part was to check whether the steady state with the periodically oscillating pairing amplitude also extends into a non-integrable region of the parameter space. 

The time evolution of the  pairing amplitude following the quench is shown in Figs.~\ref{FigDynNonint} and \ref{FigDynint}. We see that for  certain values of $h_Z/\varepsilon_F$ the order parameter magnitude $|\Delta(t)|$ shows oscillations with several frequencies and  its amplitude   is not constant at long times (at least up to the longest time scales we were able to achieve with our numerics).  However,  note the striking difference between the dynamics in Fig.~\ref{FigDynNonint} and Fig.~ \ref{FigDynint}:
when $h_{Zf}$ exceeds the value of $h_{c3}\approx 1.02\varepsilon_F$ provided $h_{Zi}=1.85\varepsilon_F$, the pairing amplitude  shows regular oscillations with  constant amplitude. This behavior is characteristic of $\Delta(t)$ which is found in exactly solvable limit. 

\begin{figure}[ht]
\includegraphics[width=0.9\linewidth]{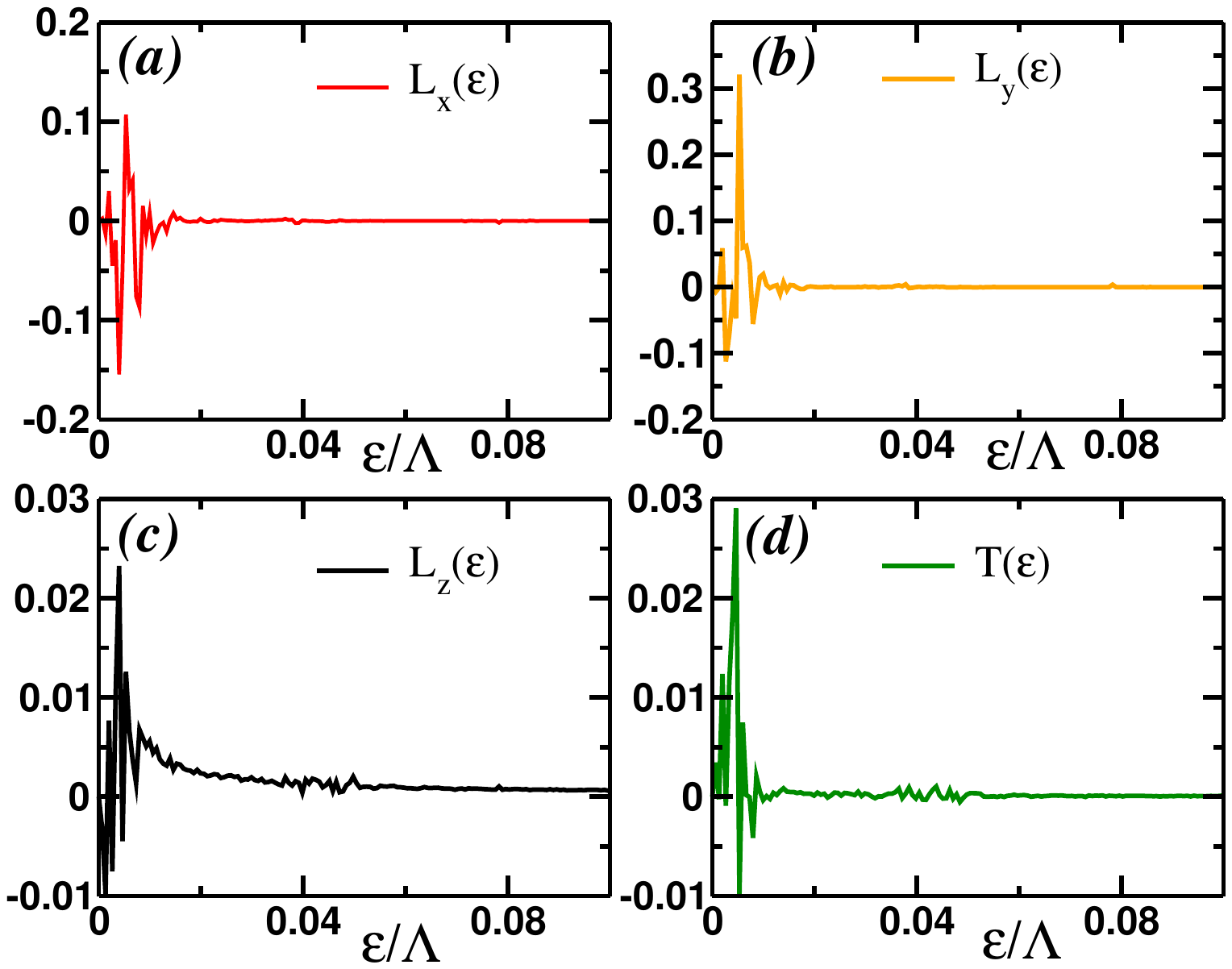}%
\caption{\label{FigDynNonintLT} (Color online)  Energy dependence of ${\vec L}(\varepsilon)$ and $T(\varepsilon)$ at time
$t\delta=2.05$ ($\delta$ is a level spacing). The values of the remaining parameters are: $h_{Zi}=1.85\varepsilon_F$, $h_{Zf}=0.5\varepsilon_F$, $n_c=0.125$,
$\Lambda=10\varepsilon_F$ and $\varepsilon_F=0.785$.}
\end{figure}

\begin{figure}[ht]
\includegraphics[width=0.9\linewidth]{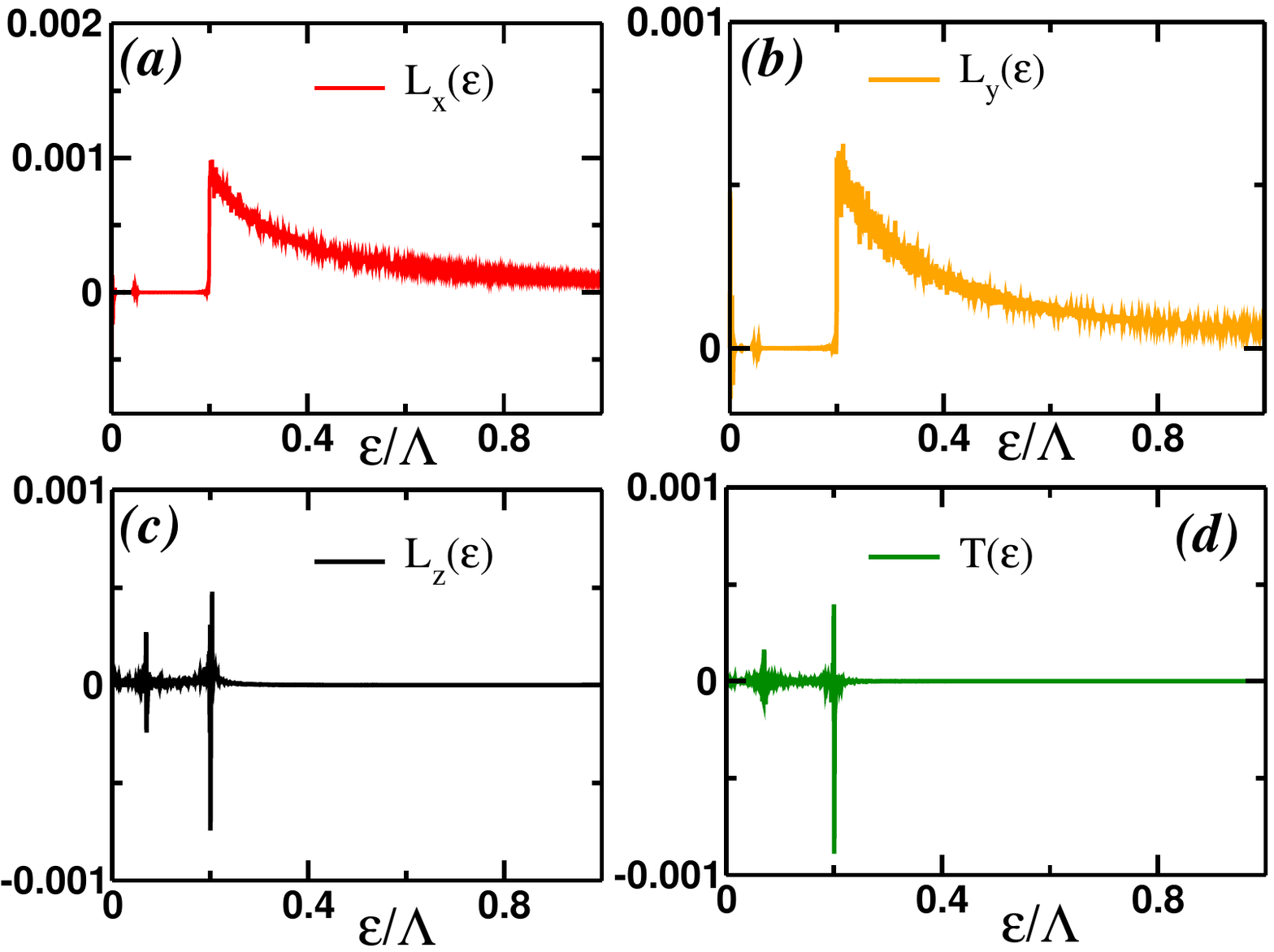}%
\caption{\label{FigDynIntLT} (Color online) Same as Fig. \ref{FigDynNonintLT} with $h_{Zf}=1.25\varepsilon_F$. In contrast with the Fig. 
\ref{FigDynNonintLT} we see that all components of ${\vec L}(\varepsilon,t\to\infty)$ as well as $T(\varepsilon,t\to\infty)$ are vanishingly small for all single particle energies. }
\end{figure}

To get further insight into the origin of this behavior, on Figs. \ref{FigDynNonintLT} and \ref{FigDynIntLT} we plot the single particle energy dependence of the auxiliary functions ${\vec L}(\varepsilon,t)$ and $T(\varepsilon,t)$ at long times when $h_{Zi}<h_{c3}$ and $h_{Zi}>h_{c3}$. For these plots the regular oscillatory behavior of $\Delta(t)$ becomes clear since for $h_{Zi}>h_{c3}$ equations of motion for the functions ${\vec S}_{\bk\lambda}(t)$ decouple from the remaining four equations of motion (\ref{EqMot4L}) and (\ref{EqMot4T}). Lastly we make one more observation: this dynamical decoupling happens exactly when the system goes through the Floquet topological transition \cite{Matt2014PRL,Pu2015} corresponding to the transition from topologically trivial Floquet spectrum to a steady state with topologically non-trivial Floquet spectrum. 
However, the detailed analysis of this transition goes beyond the scope of this paper and we leave it for the future publication.

\section{Conclusions}
In this paper we have analyzed the far-from-equilibrium pairing dynamics of the spin-orbit coupled fermions in 2d with population imbalance. Specifically, we have considered two separate cases. In the first case the dynamics is initiated by the a sudden change of the pairing strength. We found that the steady state with periodically varying pairing amplitude is realized in much narrow regions of the steady-state phase diagram compared to what happens when 
spin-orbit coupling is zero. 

Exact integrability of the problem with zero imbalance implies that we can also provide analytical description for the dynamics initiated by a sudden change of imbalance to zero. We find that when the initial value of the imbalance field $h_Z$ exceeds some critical value $h_{c2}$, the steady state with periodically oscillating pairing amplitude is realized
and determine an analytical expression for $\Delta(t)$. 

Perhaps our most interesting result is our finding of the dynamical decoupling for the quenches to finite values of the population imbalance. Specifically, our numerical analysis of the equations of motion showed that when final value of the population imbalance exceeds some value $h_t$, the pairing amplitude is determined by a reduced number of the "pseudospin" variables. Interestingly, the value of the $h_t$ is a critical value separating the regions
of topologically trivial and topologically non-trivial Floquet spectrum.\cite{Pu2015} Whether topology plays a defining role in the above mentioned reduction or it is just a mere coincidence is an exciting issue, which we leave for the future studies. \\

\noindent{\bf Note added:} The steady-state diagram shown in Fig. 2 is incorrect and the pseudospin equation of motion (2.7) is missing a term. See Fig. 18 and Eq. (2.11) in Phys. Rev. B 99, 054520 (2019) for the correct steady-state diagram and equations of motion. 

\section{Acknowledgements} M.D. thanks Instituto Superior Tecnico (Lisbon, Portugal) where  part of this work was done for hospitality and partial financial support by Fund\~{a}cao para a Ci\^{e}ncia e a Tecnologia, Grant No. PTDC/FIS/111348/2009. We thank Boris Altshuler, Antonio Garcia-Garcia, Pedro Ribeiro, Pedro Sacramento and, especially, Matthew Foster, for illuminating discussions. We also would like to thank Mubarak AlQahtani for the collaboration during the initial stages of this project. This work was financially supported in part by the David and Lucile Packard Foundation (E. A.Y.),  NSF Grant No. DMR-1506547 (A. K. and M. D.) and MPI-PKS (M. D.)

\appendix
\section{Equations of motion for the single particle correlators.}
In this section we will analyze the ground state properties of the Hamiltonian (\ref{HmfChiral}) using the equations of motion of the single particle correlators. The main idea is to derive the set of the self-consistent equations describing the collisionless evolution of the pairing amplitude. 

Consider the equations of motion for the fermionic operators. We have
\beg\label{EqMota}
\begin{split}
i\frac{\partial}{\partial t}\hat{a}_{\bk\lambda}(t)&=\varepsilon_{\bk\lambda}\hat{a}_{\bk\lambda}-
\eta_\bk^*\Delta\left[\lambda\Theta_k\hat{a}_{-\bp\lambda}\dg+\widetilde{\Theta}_k\hat{a}_{-\bp\overline{\lambda}}\dg\right], \\
i\frac{\partial}{\partial t}\hat{a}_{\bk\lambda}\dg(t)&=-\varepsilon_{\bp\lambda}\hat{a}_{\bk\lambda}\dg+
\eta_\bk\overline{\Delta}\left[\lambda\Theta_k\hat{a}_{-\bk\lambda}+\widetilde{\Theta}_k\hat{a}_{-\bk\overline{\lambda}}\right].
\end{split}
\en
Next we introduce the following correlation functions which are diagonal in new basis:
\beg\label{correlators}
\begin{split}
&G_{\bp\lambda}(t_1,t_2)=-i\left\langle\hat{T}\left(\hat{a}_{\bp\lambda}(t_1)\hat{a}_{\bp\lambda}\dg(t_2)\right)\right\rangle, \\
&F_{\bp\lambda}(t_1,t_2)=-i\lambda\eta_\bp\left\langle\hat{T}\left(\hat{a}_{\bp\lambda}(t_1)\hat{a}_{-\bp\lambda}(t_2)\right)\right\rangle, \\
&\widetilde{G}_{\bp\lambda}(t_1,t_2)=-i\left\langle\hat{T}\left(\hat{a}_{-\bp\lambda}\dg(t_1)\hat{a}_{-\bp\lambda}(t_2)\right)\right\rangle, \\
&\overline{F}_{\bp\lambda}(t_1,t_2)=-i\lambda\eta_\bp^*\left\langle\hat{T}\left(\hat{a}_{-\bp\lambda}\dg(t_1)\hat{a}_{\bp\lambda}\dg(t_2)\right)\right\rangle
\end{split}
\en
Similarly, we introduce the "off-diagonal" correlators  which account for the scattering of fermions between the two chiral bands:
\beg\label{correlators2}
\begin{split}
&\Gamma_{\bp\lambda}(t_1,t_2)=-i\lambda\left\langle\hat{T}\left(\hat{a}_{\bp\overline{\lambda}}(t_1)\hat{a}_{\bp\lambda}\dg(t_2)\right)\right\rangle, \\
&\Phi_{\bp\overline{\lambda}}(t_1,t_2)=-i\eta_\bp\left\langle\hat{T}\left(\hat{a}_{\bp{\lambda}}(t_1)\hat{a}_{-\bp\overline{\lambda}}(t_2)\right)\right\rangle, \\
&\widetilde{\Gamma}_{\bp\lambda}(t_1,t_2)=-i\lambda\left\langle\hat{T}\left(\hat{a}_{-\bp\overline{\lambda}}\dg(t_1)\hat{a}_{-\bp\lambda}(t_2)\right)\right\rangle, \\ 
&\overline{\Phi}_{\bp\lambda}(t_1,t_2)=-i\eta_\bp^*\left\langle\hat{T}\left(\hat{a}_{-\bp\overline{\lambda}}\dg(t_1)\hat{a}_{\bp\lambda}\dg(t_2)\right)\right\rangle.
\end{split}
\en
As a next step one can derive the equations of motion for these correlation functions using (\ref{EqMota}). 
\paragraph{Equations of motion for the diagonal chiral correlators}
For the diagonal in $\lambda$ correlation functions above we have
\beg
\begin{split}
&\left(i\frac{\partial}{\partial t_1}-\varepsilon_{\bk\lambda}\right)G_{\bk\lambda}(t_1,t_2)\\&+\Delta\left[\Theta_k\overline{F}_{\bk\lambda}(t_1,t_2)+\widetilde{\Theta}_k\overline{\Phi}_{\bk\lambda}(t_1,t_2)\right]=\delta(t_1-t_2), \\
&\left(i\frac{\partial}{\partial t_1}-\varepsilon_{\bk\overline{\lambda}}\right)\Gamma_{\bk\lambda}(t_1,t_2)\\&-\Delta\left[\Theta_k\overline{\Phi}_{\bk\lambda}(t_1,t_2)-\widetilde{\Theta}_k\overline{F}_{\bk\lambda}(t_1,t_2)\right]=0, \\
\end{split}
\en
Similarly, the equations of motion for the anomalous correlation functions (\ref{correlators2}) are:
\beg
\begin{split}
&\left(i\frac{\partial}{\partial t_1}+\varepsilon_{\bk\lambda}\right)\overline{F}_{\bk\lambda}(t_1,t_2)\\&+\overline{\Delta}\left[\Theta_kG_{\bk\lambda}(t_1,t_2)+\widetilde{\Theta}_k{\Gamma}_{\bk\lambda}(t_1,t_2)\right]=0, \\
&\left(i\frac{\partial}{\partial t_1}+\varepsilon_{\bk\overline{\lambda}}\right)\overline{\Phi}_{\bk\lambda}(t_1,t_2)\\&-\overline{\Delta}\left[\Theta_k\Gamma_{\bk\lambda}(t_1,t_2)-\widetilde{\Theta}_k{G}_{\bk\lambda}(t_1,t_2)\right]=0, \\
&\left(i\frac{\partial}{\partial t_1}+\varepsilon_{\bk\overline{\lambda}}\right)\widetilde{\Gamma}_{\bk\lambda}(t_1,t_2)\\&-\overline{\Delta}\left[\Theta_k{\Phi}_{\bk\lambda}(t_1,t_2)-\widetilde{\Theta}_k{F}_{\bk\lambda}(t_1,t_2)\right]=0.
\end{split}
\en
In equilibrium, all these correlation functions depend on $(t_1-t_2)$ only, so we can perform the Fourier transform and compute them explicitly. It follows:
\beg
\begin{split}
&\Gamma_{\bk\lambda}(\omega)=\frac{(\omega+\varepsilon_{\bk\overline{\lambda}})\Delta\widetilde{\Theta}_k\overline{F}_{\bk\lambda}(\omega)-\Delta^2\Theta_k\widetilde{\Theta}_kG_{\bk\lambda}(\omega)}{\omega^2-\varepsilon_{\bk\overline{\lambda}}^2-\Delta^2\Theta_k^2}, \\
&\overline{\Phi}_{\bk\lambda}(\omega)=-\frac{(\omega-\varepsilon_{\bk\overline{\lambda}})\overline{\Delta}\widetilde{\Theta}_kG_{\bk\lambda}(\omega)+\Delta^2\Theta_k\widetilde{\Theta}_k\overline{F}_{\bk\lambda}(\omega)}{\omega^2-\varepsilon_{\bk\overline{\lambda}}^2-\Delta^2\Theta_k^2}.
\end{split}
\en
where we assume that $\Delta=\overline{\Delta}$, i.e. in the ground state the pairing amplitude is real. We have
\beg\label{GammaEQ}
\begin{split}
G_{\bk\lambda}(\omega)&=-\frac{(\omega+\varepsilon_{\bk\lambda})[\Theta_k^2\Delta^2+\varepsilon_{\bk\overline{\lambda}}^2-\omega^2]}{(\omega^2-E_{\bk\lambda}^2)(\omega^2-E_{\bk\overline{\lambda}}^2)}\\&+\frac{\widetilde{\Theta}_k^2\Delta^2(\omega+\varepsilon_{\bk\overline{\lambda}})}{(\omega^2-E_{\bk\lambda}^2)(\omega^2-E_{\bk\overline{\lambda}}^2)}, \\
\overline{F}_{\bk\lambda}(\omega)&=\frac{\Theta_k\Delta[\Delta^2+\varepsilon_{\bk\overline{\lambda}}^2-\omega^2]}{(\omega^2-E_{\bk\lambda}^2)(\omega^2-E_{\bk\overline{\lambda}}^2)}, \\ {\Gamma}_{\bk\lambda}(\omega)&=\frac{2\lambda\Theta_k\widetilde{\Theta}_k\Delta^2R_k}{(\omega^2-E_{\bk\lambda}^2)(\omega^2-E_{\bk\overline{\lambda}}^2)}, \\
\overline{\Phi}_{\bk\lambda}(\omega)&=\frac{\widetilde{\Theta}_k\Delta[\Delta^2+(\varepsilon_{\bk\overline{\lambda}}-\omega)(\omega+\varepsilon_{\bk{\lambda}})]}{(\omega^2-E_{\bk\lambda}^2)(\omega^2-E_{\bk\overline{\lambda}}^2)},
\\ \widetilde{\Gamma}_{\bk\lambda}(\omega)&=-\frac{2\lambda\Theta_k\widetilde{\Theta}_k\Delta^2R_k}{(\omega^2-E_{\bk\lambda}^2)(\omega^2-E_{\bk\overline{\lambda}}^2)}
\end{split}
\en
The last expression follows from the symmetry properties of the corresponding equations of motion. Note that from these expressions it follows
\beg
\begin{split}
&\overline{\Phi}_{\bk\lambda}(\omega)=\overline{\Phi}_{\bk\overline{\lambda}}(-\omega), \quad
{\Gamma}_{\bk\lambda}(\omega)=-{\Gamma}_{\bk\overline{\lambda}}(\omega), \\ 
&\widetilde{\Gamma}_{\bk\lambda}(\omega)={\Gamma}_{\bk\overline{\lambda}}(\omega).
\end{split}
\en

To compute the averages which enter into the self-consistency equation which determines $\Delta$, we employthe  Matsubara frequency representation $\omega\to i\omega_n$. Then performing the summations over the Matsubara
frequencies and take the limit $T\to 0$. The resulting functions of momentum are listed in the main text, Eqs. (\ref{SxLxInit},\ref{SzLzInit}). Note that $L^z\propto\langle{a_{\bk\overline{\lambda}}\dg a_{\bk\lambda}}\rangle$ is generated already within the mean-field theory
despite the fact that the terms proportional to $a_{\bk\overline{\lambda}}\dg a_{\bk\lambda}$ do not enter into the Hamiltonian.
In what follows, we will also consider function
\beg
T_{\bk}^z=\frac{1}{2}T\sum\limits_{i\omega_n}\left[{{\Gamma}_{\bk\overline{\lambda}}(i\omega_n)+{\Gamma}_{\bk\lambda}(i\omega_n)}\right]
\en
which is zero in the ground state, however it is generated during the evolution. 

Our goal now is to derive the equations of motion for all the correlation functions above as a function of 
\beg
t=\frac{t_1+t_2}{2}.
\en
Since both normal and anomalous correlators (\ref{correlators},\ref{correlators2})
depend on $\tau=t_1-t_2$, but the order parameter $\Delta(t)$ is a function of total time $t$ only. Thus, in what follows
we consider $\tau=0$.

From equations of motion for the fermionic operators (\ref{EqMota}) and (\ref{correlators},\ref{correlators2}) it follows
\beg
\begin{split}
&i\frac{d}{dt}G_{\bk\lambda}(t)+\Delta(t)\left[\Theta_k\overline{F}_{\bk\lambda}(t)+\widetilde{\Theta}_k\overline{\Phi}_{\bk\lambda}(t)\right]\\&-
\overline{\Delta}(t)\left[\Theta_k{F}_{\bk\lambda}(t)+\widetilde{\Theta}_k{\Phi}_{\bk\overline{\lambda}}(t)\right]=0, \\
&\left(i\frac{d}{dt}-2\varepsilon_{\bk\lambda}\right)F_{\bk\lambda}(t)+\Delta(t)\left\{\Theta_k\left[\widetilde{G}_{\bk\lambda}(t)-G_{\bk\lambda}(t)\right]\right.\\&\left.+\widetilde{\Theta}_k\left[\widetilde{\Gamma}_{\bk\lambda}(t)-\Gamma_{\bk\lambda}(t)\right]\right\}=0.
\end{split}
\en
Similarly for the remaining two correlation functions which are diagonal in new basis I find
\beg
\begin{split}
&i\frac{d}{dt}\widetilde{G}_{\bk\lambda}(t)+\overline{\Delta}(t)\left[\Theta_k{F}_{\bk\lambda}(t)+\widetilde{\Theta}_k{\Phi}_{\bk\overline{\lambda}}(t)\right]\\&-\Delta(t)\left[\Theta_k\overline{F}_{\bk\lambda}(t)+\widetilde{\Theta}_k\overline{\Phi}_{\bk\lambda}(t)\right]=0, \\
&\left(i\frac{d}{dt}+2\varepsilon_{\bk\lambda}\right)\overline{F}_{\bk\lambda}(t)+\overline{\Delta}(t)\left\{\Theta_k\left[G_{\bk\lambda}(t)-\widetilde{G}_{\bk\lambda}(t)\right]\right.\\&\left.+\widetilde{\Theta}_k\left[\Gamma_{\bk\lambda}(t)-\widetilde{\Gamma}_{\bk\lambda}(t)\right]\right\}=0.
\end{split}
\en
From the equations for the normal propagators it follows 
\beg\label{tGG}
\widetilde{G}_{\bk\lambda}(t)=-{G}_{\bk\lambda}(t).
\en
Let us introduce the following functions
\beg\label{SSpins}
\begin{split}
&S_{\bk\lambda}^z(t)=\frac{i}{2}\left[\tilde{G}_{\bk\lambda}(t)-{G}_{\bk\lambda}(t)\right], \\
&S_{\bk\lambda}^{-}(t)=S_{\bk\lambda}^x(t)-iS_{\bk\lambda}^y(t)=-iF_{\bk\lambda}(t), \\
&S_{\bk\lambda}^{+}(t)=S_{\bk\lambda}^x(t)+iS_{\bk\lambda}^y(t)=-i\overline{F}_{\bk\lambda}(t)
\end{split}
\en
and
\beg\label{LSpins}
\begin{split}
&L_{\bk\lambda}^z(t)=\frac{i}{2}\left[\widetilde{\Gamma}_{\bk\lambda}(t)-{\Gamma}_{\bk\lambda}(t)\right], \\
&L_{\bk\lambda}^{-}(t)=L_{\bk\lambda}^x(t)-iL_{\bk\lambda}^y(t)=-i\Phi_{\bk\overline{\lambda}}(t), \\
&L_{\bk\lambda}^{+}(t)=L_{\bk\lambda}^x(t)+iL_{\bk\lambda}^y(t)=-i\overline{\Phi}_{\bk\lambda}(t)
\end{split}
\en
In terms of these new functions, self-consistency equation for the pairing amplitude reads
\beg
\Delta(t)=g\sum\limits_{\bk\mu}\left[\Theta_k S_{\bk\mu}^{-}(t)+\widetilde{\Theta}_k L_{\bk\mu}^{-}(t)\right].
\en
\paragraph{Equations of motion for the off-diagonal chiral correlators}
The remaining equations of motion for the components of ${\vec L}$ can be derived in the same way. Let us obtain the equations
of motion for $\Gamma_{\bk\lambda}(t)$. In what follows the only relation I will use is
\beg\label{Gammakl}
\begin{split}
&\Gamma_{\bk\overline{\lambda}}(t)=\widetilde{\Gamma}_{\bk\lambda}(t), \quad \overline{\Phi}_{\bk\lambda}(t)=\overline{\Phi}_{\bk\overline{\lambda}}(t), \\ &\widetilde{G}_{\bk{\lambda}}=-{G}_{\bk\lambda}.
\end{split}
\en
The validity of these relations will be proven when we analyze equilibrium. We need to keep in mind, however, that given the second
relation we expect that equations of motion for ${L}_{\bk\lambda}^\pm(t)$ should not depend on chiral band index $\lambda$. 
The equations of motion for the correlator $\overline{\Phi}_{\bk\lambda}(t_1,t_2)$ are
\beg
\begin{split}
&\left(i\frac{\partial}{\partial t_1}+\varepsilon_{\bk\overline{\lambda}}\right)\overline{\Phi}_{\bk\lambda}(t_1,t_2)\\&-\overline{\Delta}\left[\Theta_k\Gamma_{\bk\lambda}(t_1,t_2)-\widetilde{\Theta}_k{G}_{\bk\lambda}(t_1,t_2)\right]=0, \\
&\left(i\frac{\partial}{\partial t_2}+\varepsilon_{\bk{\lambda}}\right)\overline{\Phi}_{\bk\lambda}(t_1,t_2)\\&-\overline{\Delta}\left[\Theta_k\widetilde{\Gamma}_{\bk\lambda}(t_1,t_2)+\widetilde{\Theta}_k\widetilde{G}_{\bk\overline{\lambda}}(t_1,t_2)\right]=0.
\end{split}
\en
where we used $\eta_{-\bp}=-\eta_{\bp}$. 
Adding these two equations yields
\beg
\left(i\frac{\partial}{\partial t}+2\epsilon_{\bk}\right)\overline{\Phi}_{\bk\lambda}(t)-
\widetilde{\Theta}_k\overline{\Delta}(t)\left[\widetilde{G}_{\bk\overline{\lambda}}(t)-G_{\bk\lambda}(t)\right]=0
\en
From this equation we can immediately obtain the equations of motion for $L_{\bk\lambda}^{x,y}$ using Eqs. (\ref{SSpins},\ref{LSpins}).

Lastly, we derive the equation of motion for 
\beg
T_\bk(t)=\frac{\Gamma_{\bk\lambda}(t)+\Gamma_{\bk\overline{\lambda}}(t)}{2}.
\en
Before I write down this equation, let me first obtain the equations of motion for $\Gamma_{\bk\lambda}$ and $\widetilde{\Gamma}_{\bk\lambda}$. We have:
\beg
\begin{split}
&\left(i\frac{\partial}{\partial t_1}-\varepsilon_{\bk\overline{\lambda}}\right)\Gamma_{\bk\lambda}(t_1,t_2)\\&-\Delta\left[\Theta_k\overline{\Phi}_{\bk\lambda}(t_1,t_2)-\widetilde{\Theta}_k\overline{F}_{\bk\lambda}(t_1,t_2)\right]=0, \\
&\left(i\frac{\partial}{\partial t_2}+\varepsilon_{\bk{\lambda}}\right){\Gamma}_{\bk\lambda}(t_1,t_2)\\&-\overline{\Delta}\left[\Theta_k{\Phi}_{\bk\lambda}(t_1,t_2)-\widetilde{\Theta}_k{F}_{\bk\overline{\lambda}}(t_1,t_2)\right]=0, \\
&\left(i\frac{\partial}{\partial t_1}+\varepsilon_{\bk\overline{\lambda}}\right)\widetilde{\Gamma}_{\bk\lambda}(t_1,t_2)\\&-\overline{\Delta}\left[\Theta_k{\Phi}_{\bk\lambda}(t_1,t_2)-\widetilde{\Theta}_k{F}_{\bk\lambda}(t_1,t_2)\right]=0, \\
&\left(i\frac{\partial}{\partial t_2}-\varepsilon_{\bk{\lambda}}\right)\widetilde{\Gamma}_{\bk\lambda}(t_1,t_2)\\&-{\Delta}\left[\Theta_k\overline{\Phi}_{\bk\lambda}(t_1,t_2)-\widetilde{\Theta}_k\overline{F}_{\bk\overline{\lambda}}(t_1,t_2)\right]=0.
\end{split}
\en
where we have employed (\ref{Gammakl}), $\eta_{-\bp}=-\eta_{\bp}$ and $\lambda\overline{\lambda}=-1$. Adding the first and the second equations 
and then the third and the fourth one and setting $\tau=t_1-t_2=0$ yields
\beg\label{EOMGammas}
\begin{split}
&\left(i\frac{\partial}{\partial t}-2\lambda R_k\right)\Gamma_{\bk\lambda}(t)-\Theta_k\left[\Delta\overline{\Phi}_{\bk\lambda}(t)+
\overline{\Delta}{\Phi}_{\bk\lambda}(t)\right]\\&+\widetilde{\Theta}_k\left[\Delta
\overline{F}_{\bk\lambda}(t)+\overline{\Delta}{F}_{\bk\overline{\lambda}}(t)\right]=0, \\
&\left(i\frac{\partial}{\partial t}+2\lambda R_k\right)\widetilde{\Gamma}_{\bk\lambda}(t)-\Theta_k\left[\Delta\overline{\Phi}_{\bk\lambda}(t)+
\overline{\Delta}{\Phi}_{\bk\lambda}(t)\right]\\&+\widetilde{\Theta}_k\left[\overline{\Delta}{F}_{\bk{\lambda}}(t)+\Delta
\overline{F}_{\bk\overline{\lambda}}(t)\right]=0.
\end{split}
\en 
where $R_k=\sqrt{V_z^2+\alpha_{SO}^2k^2}$.
From these equations we see that given the property (\ref{Gammakl}) we have
\beg
\Gamma_{\bk\lambda}=Z_\bk+iL_{\bk\lambda}^z.
\en
It is now straightforward to verify that the equations of motion for these objects are the same as the ones listed in the main text, Eqs. (\ref{EqMot4S}, \ref{EqMot4L}, \ref{EqMot4T}).
Thus we have ten equations of motion. These equations are decoupled into six plus four when either $\alpha_{SO}=0$ or $h_Z=0$.Note that both $L_z$ and $T_\bk$ do not enter into the Hamiltonian and are generated in the course of dynamics.

\section{general relations between the components auxiliary functions in equilibrium}

We assume that in equilibrium $\Delta_x=\Delta$ and $\Delta_y=0$. This implies that both at $t=0$ both 
$S_{\bk\lambda}^y=0$ and $L_{\bk\lambda}^y=0$ in accordance with the self-consistency conditions.
This guarantees that seven out of ten equations (\ref{EqMot4S},\ref{EqMot4L},\ref{EqMot4T}) for the components of vectors ${\vec S}$, ${\vec L}$ and $T_\bk$ are identically zero. The remaining three equations are
\beg\label{Check1}
\begin{split}
&\varepsilon_{\bk\lambda}S_{\bk\lambda}^x+\Delta\cdot\left(
\Theta_kS_{\bk\lambda}^z+\widetilde{\Theta}_kL_{\bk\lambda}^z\right)=0, \\ &2\epsilon_\bk L_{\bk\lambda}^x+\widetilde{\Theta}_k\Delta\left[S_{\bk\lambda}^z+S_{\bk\overline{\lambda}}^z\right]=0, \\
&2\lambda R_kL_{\bk\lambda}^z+\Delta\left\{2\Theta_kL_{\bk\lambda}^x
-\widetilde{\Theta}_k\left[S_{\bk\lambda}^x+S_{\bk\overline{\lambda}}^x\right]\right\}=0.
\end{split}
\en
Lastly, let us verify if expressions for the spin components satisfy (\ref{Check1}). For the first two equations we find:
\beg
\begin{split}
&\Delta\left[\Theta_kS_{\bk\lambda}^z+\widetilde{\Theta}_kL_{\bk\lambda}^z\right]=-\varepsilon_{\bk\lambda}S_{\bk\lambda}^x, \\
&\widetilde{\Theta}_k\Delta\left[S_{\bk\lambda}^z+S_{\bk\overline{\lambda}}^z\right]=-2\epsilon_\bk L_{\bk\lambda}^x.
\end{split}
\en
Lastly, let us check the third equation (\ref{Check1}):
\beg
\begin{split}
&2\Theta_k\Delta L_{\bk\lambda}^x
-\widetilde{\Theta}_k\Delta\left[S_{\bk\lambda}^x+S_{\bk\overline{\lambda}}^x\right]\\&=-\frac{4\Theta_k\widetilde{\Theta}_k\Delta^2R_k^2}{2E_{\bk\lambda}E_{\bk\overline{\lambda}}(E_{\bk\lambda}+E_{\bk\overline{\lambda}})}
\end{split}
\en
On the other hand
\beg
\begin{split}
2\lambda R_k L_{\bk\lambda}^z&=2\lambda R_k\frac{\Theta_k\widetilde{\Theta}_k\Delta^2(\varepsilon_{\bk\overline{\lambda}}-\varepsilon_{\bk\lambda})}{2E_{\bk\lambda}E_{\bk\overline{\lambda}}(E_{\bk\lambda}+E_{\bk\overline{\lambda}})}\\&=\frac{4\Theta_k\widetilde{\Theta}_k\Delta^2R_k^2}{2E_{\bk\lambda}E_{\bk\overline{\lambda}}(E_{\bk\lambda}+E_{\bk\overline{\lambda}})}.
\end{split}
\en
Thus the third equation in (\ref{Check1}) holds. 

\bibliography{RefsQuench2dSO}

\end{document}